\documentclass[letterpaper,twocolumn,10pt]{article}
\usepackage{ligroup}
\usepackage{xspace}  
\usepackage{pifont}
\usepackage{placeins}
\usepackage[absolute]{textpos}

\usepackage{amsmath}
\usepackage{amssymb}

\usepackage[table]{xcolor}
\definecolor{aprowcolor}{gray}{0.9} 
\definecolor{lightgray}{gray}{0.94}
\definecolor{bordergray}{gray}{0.7}

\usepackage{tikz}
\usepackage[most]{tcolorbox}
\usepackage[framemethod=TikZ]{mdframed}

\newcommand*\filledcircled[2][\normalsize]{%
  \tikz[baseline=(char.base)]{
    \node[shape=circle,fill,inner sep=0.5pt] (char) {#1\textcolor{white}{#2}};}}

\usepackage{algorithmic}
\usepackage{algorithm}
\usepackage{textcomp}
\usepackage{stfloats}
\usepackage{verbatim}
\usepackage{graphicx}
\usepackage{multirow}  
\usepackage{booktabs} 

\newcommand{\mypara}[1]{\noindent\textbf{\textit{#1}.}\xspace}

\newmdenv[
  backgroundcolor=lightgray,
  linecolor=bordergray,
  linewidth=0.8pt,
  roundcorner=4pt,
  innertopmargin=6pt,
  innerbottommargin=6pt,
  innerleftmargin=8pt,
  innerrightmargin=8pt,
  skipabove=10pt,
  skipbelow=10pt
]{remarkbox}

\AtBeginDocument{
  
}

\usepackage[hyphens]{xurl}   
\usepackage{hyperref}
\hypersetup{
  colorlinks,
  urlcolor={orange!70!red}
}

\begin{document}

\date{}

\title{\bf Beyond Known Fakes: Generalized Detection of AI-Generated Images via Post-hoc Distribution Alignment}

\author{
Li Wang\textsuperscript{1,3,4}, 
Wenyu Chen\textsuperscript{1}, 
Xiangtao Meng\textsuperscript{1}, 
Zheng Li\textsuperscript{1,3,4}\thanks{Corresponding authors}, 
Shanqing Guo\textsuperscript{1,3,4}\footnotemark[1]
\\[1ex]
\textsuperscript{1}\textit{School of Cyber Science and Technology, Shandong University} \\
\textsuperscript{3}\textit{State Key Laboratory of Cryptography and Digital Economy Security, Shandong University} \\
\textsuperscript{4}\textit{Shandong Key Laboratory of Artificial Intelligence Security, Shandong University}
}

\maketitle

\begin{abstract}
The rapid proliferation of highly realistic AI-generated images poses serious security threats such as misinformation and identity fraud. Detecting generated images in open-world settings is particularly challenging when they originate from unknown generators, as existing methods typically rely on model-specific artifacts and require retraining on new fake data, limiting their generalization and scalability. 
In this work, we propose \textit{Post-hoc Distribution Alignment (PDA)}, a generalized and model-agnostic framework for detecting AI-generated images under unknown generative threats. 
Specifically, PDA reformulates detection as a distribution alignment task by regenerating test images through a known generative model. 
When real images are regenerated, they inherit model-specific artifacts and align with the known fake distribution. In contrast, regenerated unknown fakes contain incompatible or mixed artifacts and remain misaligned. 
This difference allows an existing detector, trained on the known generative model, to accurately distinguish real images from unknown fakes without requiring access to unseen data or retraining. 
Extensive experiments across 16 state-of-the-art generative models, including GANs, diffusion models, and commercial text-to-image APIs (e.g., Midjourney), demonstrate that PDA achieves average detection accuracy of 96.69\%, outperforming the best baseline by 10.71\%. Comprehensive ablation studies and robustness analyses further confirm PDA’s generalizability and resilience to distribution shifts and image transformations. Overall, our work provides a practical and scalable solution for real-world AI-generated image detection where new generative models emerge continuously. 
\end{abstract}

\section{Introduction} 
\label{intro}
Recent advances in generative models have fueled the rapid proliferation of AI-generated images~\cite{wang2024dp,sha2023fake,zheng2024breaking,wang2024moderator,zhang2025detecting,gu2022vector}, enabling the mass creation of highly realistic synthetic visual content. 
While these technologies unlock broad applications in design, entertainment, and virtual reality, they simultaneously introduce growing security and privacy risks~\cite{abdullah2024analysis,li2022seeing,meng2024ava,tang2025towards}. 
Malicious actors can exploit generative tools to fabricate deceptive imagery or impersonate identities, thereby undermining content authenticity and eroding trust in digital information ecosystems~\cite{barrett2023identifying,cao2023impress}. 
These risks have already manifested in real-world incidents. 
For example, non-consensual AI-generated explicit images of public figures have circulated widely (e.g., reported by \textit{CBS News}\footnote{\url{https://www.cbsnews.com/news/taylor-swift-deepfakes-online-outrage-artificial-intelligence/}}
), and fabricated AI-generated images were used to spread false narratives about the 2025 Bondi Beach shooting, leading to large-scale public misinformation\footnote{\url{https://www.abc.net.au/news/verify-disinformation-and-deepfakes-after-bondi-attack/106154250}}. 
Such incidents are further exacerbated by the increasing accessibility and realism of modern generative systems~\cite{li2025artificial,midjourney_website}. 
As generative models continue to evolve rapidly and malicious actors can readily adopt new and diverse generators, a central security challenge emerges: developing AI-generated image detectors that remain reliable in open-world settings, where test samples may originate from previously unknown generative models~\cite{meng2024ava,an2024rethinking,deng2024towards}. 


\begin{figure}[!t]
  \centering
  \includegraphics[width=0.98\linewidth]{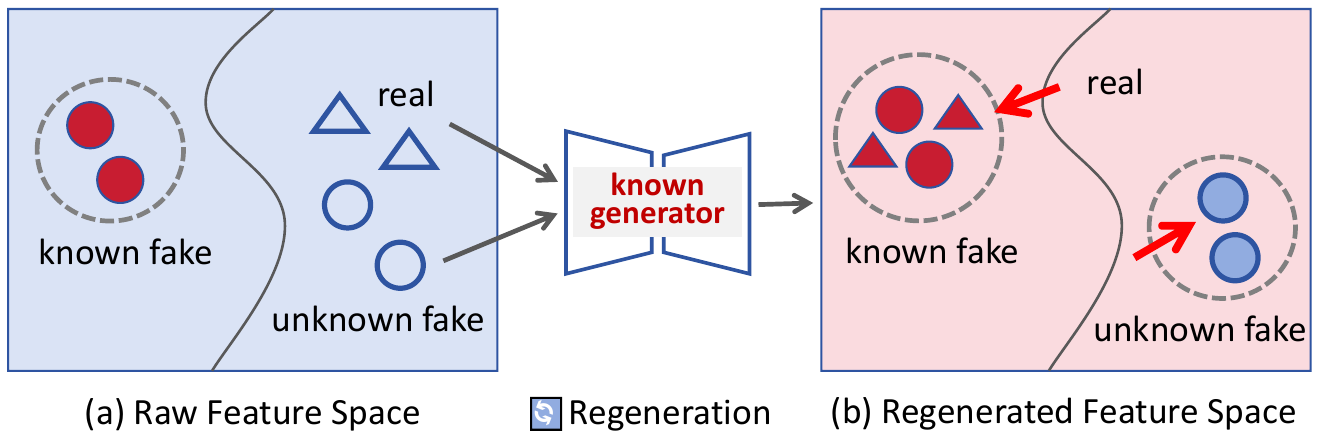}
 \caption{High-level illustration of PDA: Reals become distributionally aligned with known fakes through regeneration, while unknown fakes remain misaligned in the feature space.}
\label{fig:illustration}
\end{figure}

Despite growing research efforts~\cite{sha2024zerofake,ojha2023towards,chen2024drct}, existing AI-generated image detection paradigms suffer from a fundamental limitation: \textit{poor generalization to fake images generated by unknown generative models}. Most current detectors are trained on fake images from a limited set of known generators, relying heavily on model-specific artifacts~\cite{zheng2024breaking,corvi2023intriguing,wang2023dire,wang2020cnn}. As illustrated in \autoref{fig:illustration} (a), these approaches learn clear decision boundaries that effectively separate known fakes from real images.
However, modern generative models differ substantially in architecture, training data, and rendering pipelines, resulting in highly diverse and generator-dependent artifacts that do not transfer reliably across models~\cite{liang2024comprehensive,ojha2023towards}. 
Under this distribution shift, unknown fakes often overlap with real images in the learned feature space, leading to systematic misclassification and undermining detection reliability in realistic open-world deployments~\cite{lin2025conformal,yu2025learning}. 
A common mitigation strategy is to continuously retrain or fine-tune detectors as new generators emerge~\cite{epstein2023online,chen2022ost,tang2025towards}. However, retraining-centric strategies assume timely access to labeled fake data and incurs substantial computational overhead, making it impractical in latency-sensitive and rapidly evolving adversarial environments.


To address these challenges, we propose \textit{Post-hoc Distribution Alignment (PDA)}, a novel framework for generalized detection of AI-generated images under unknown generative threats. Instead of modeling the diverse and evolving distributions of \textit{fake images}, PDA shifts the detection focus to \textit{real images}: by regenerating real images through a single known generative model, they inherit consistent model-specific artifacts and thus align with the known fake distribution. In contrast, fake images produced by unknown generators already contain generator-dependent traces; when regenerated, these traces interfere with the artifacts introduced by the known model, resulting in hybrid or misaligned patterns and inducing distributional shifts (see \autoref{fig:illustration}(b)). 
This discrepancy enables existing detectors—trained solely on pure artifacts from the known generator—to reliably distinguish real images from unknown fakes without requiring access to unseen generators or additional retraining. 
Concretely, PDA adopts a three-step detection strategy (see~\autoref{three-step detection}). 
\textbf{Step \filledcircled[\small]{1} Raw-space Filtering:} PDA first evaluates whether a test image aligns with the known fake distribution in the raw feature space, defined by an existing detector. 
If not—indicating a real or unknown fake—it proceeds to the next step.
\textbf{Step \filledcircled[\small]{2} Regeneration}: The image is then regenerated using a known generative model to produce a \textit{pseudo-fake} version. 
Real images transformed in this way align with known fake distributions, whereas unknown fakes remain misaligned. 
\textbf{Step \filledcircled[\small]{3} Differentiation}: A threshold-based criterion is then used to distinguish real images from unknown fakes in the regenerated feature space. 
This approach can effectively detect unknown fake images using \textit{only one known generative model}, making it adaptable and scalable for real-world applications.

Extensive experiments on 16 representative generative models—including GANs, diffusion models, and proprietary text-to-image systems—across two benchmark datasets, GenImage~\cite{zhu2024genimage} and AIGCDetect~\cite{zhong2023patchcraft},  demonstrate that PDA achieves consistently superior performance. Specifically, PDA achieves an average accuracy of 96.69\%, outperforming the strongest baseline (DRCT~\cite{chen2024drct}: 85.98\%) by a margin of +10.71\%. For example, on VQDM~\cite{gu2022vector}—a challenging unseen diffusion model—PDA achieves \textbf{97.87\%} accuracy, whereas ZeroFake~\cite{sha2024zerofake}, a prompt-aware method tailored to text-to-image diffusion models, reaches only \textbf{69.38\%}, resulting in a gap of over 28\%. 
These results demonstrate that PDA generalizes effectively to unknown generative models in open-world settings, without requiring retraining, prompt engineering, or access to internal generator parameters.  
We further conduct ablation studies and in-depth analyses, confirming that PDA provides a robust and generalizable defense against AI-generated image threats under continuously evolving generative models.

The main contributions are summarized as follows:  
\begin{itemize} 

\item We propose PDA, a novel detection framework that reformulates AI-generated image detection under open-world threats as a post-hoc distribution alignment problem. By aligning real images to the artifact distribution of a single known generator, PDA enables reliable detection of synthetic images from previously unknown generators, without retraining or access to unseen data. 

\item We develop a principled three-step detection strategy: (i) raw-space filtering to identify known fakes, (ii) regeneration through a known generator to induce alignment for real images but not for unknown fakes, and (iii) differentiating real images from unknown fakes using a threshold-based criterion in the regenerated feature space. This design enables generalized detection at inference time through a training-free mechanism that is agnostic to semantic priors and generator-specific assumptions. 

\item We conduct extensive evaluations on 16 diverse generative models spanning GANs, diffusion models, and commercial text-to-image systems across two benchmark datasets. PDA achieves an average detection accuracy of 96.69\%, outperforming strong baselines by an average margin of 10.71\%. Further ablation studies and analyses confirm the robustness and generalization of PDA in realistic open-world scenarios. 

\end{itemize}

\section{Preliminaries and Related Works}
\subsection{Evolution of AI-generated Images} 
The rapid advancement of generative models has significantly improved the fidelity and diversity of AI-generated images~\cite{wang2024dp,wang2024moderator,park2019semantic,podell2023sdxl}. 
Early approaches, such as Variational Autoencoders (VAEs)~\cite{kingma2013auto}, focused on learning latent representations of data distributions. 
Subsequently, Generative Adversarial Networks (GANs)~\cite{goodfellow2014generative} introduced adversarial training, enabling high-fidelity image synthesis. Later models, including BigGAN~\cite{brock2018large} and StyleGAN~\cite{karras2019style}, further improved realism and controllability. 
More recently, diffusion models~\cite{ho2020denoising} have emerged as a powerful alternative to GANs, achieving state-of-the-art performance in image generation by iteratively denoising a random distribution to produce highly detailed and diverse images~\cite{corvi2023detection}. 
Additionally, autoregressive models such as DALL·E~\cite{ramesh2022hierarchical}, along with diffusion-based models like Stable Diffusion 1.4~\cite{rombach2022high}, GLIDE~\cite{nichol2021glide}, and ADM~\cite{dhariwal2021diffusion}, have further pushed the boundaries by enabling highly realistic text-to-image generation. These developments have significantly broadened the applicability of AI-generated images across a variety of domains, including creative content generation, artistic design, and virtual reality ~\cite{gu2022vector,ricker2022towards}. 

However, the increasing realism of AI-generated images has also intensified concerns regarding their potential misuse, such as creating deepfakes or spreading misinformation. This growing realism has motivated research on AI-generated image detection that aims to generalize beyond specific generators. Nevertheless, as different generative models produce heterogeneous and often incompatible artifacts, fake images from unknown generators induce distribution shifts that cause detectors trained on limited generators to struggle in open-world settings. 


\begin{figure*}[t!]
  \centering
  \includegraphics[width=0.98\linewidth]{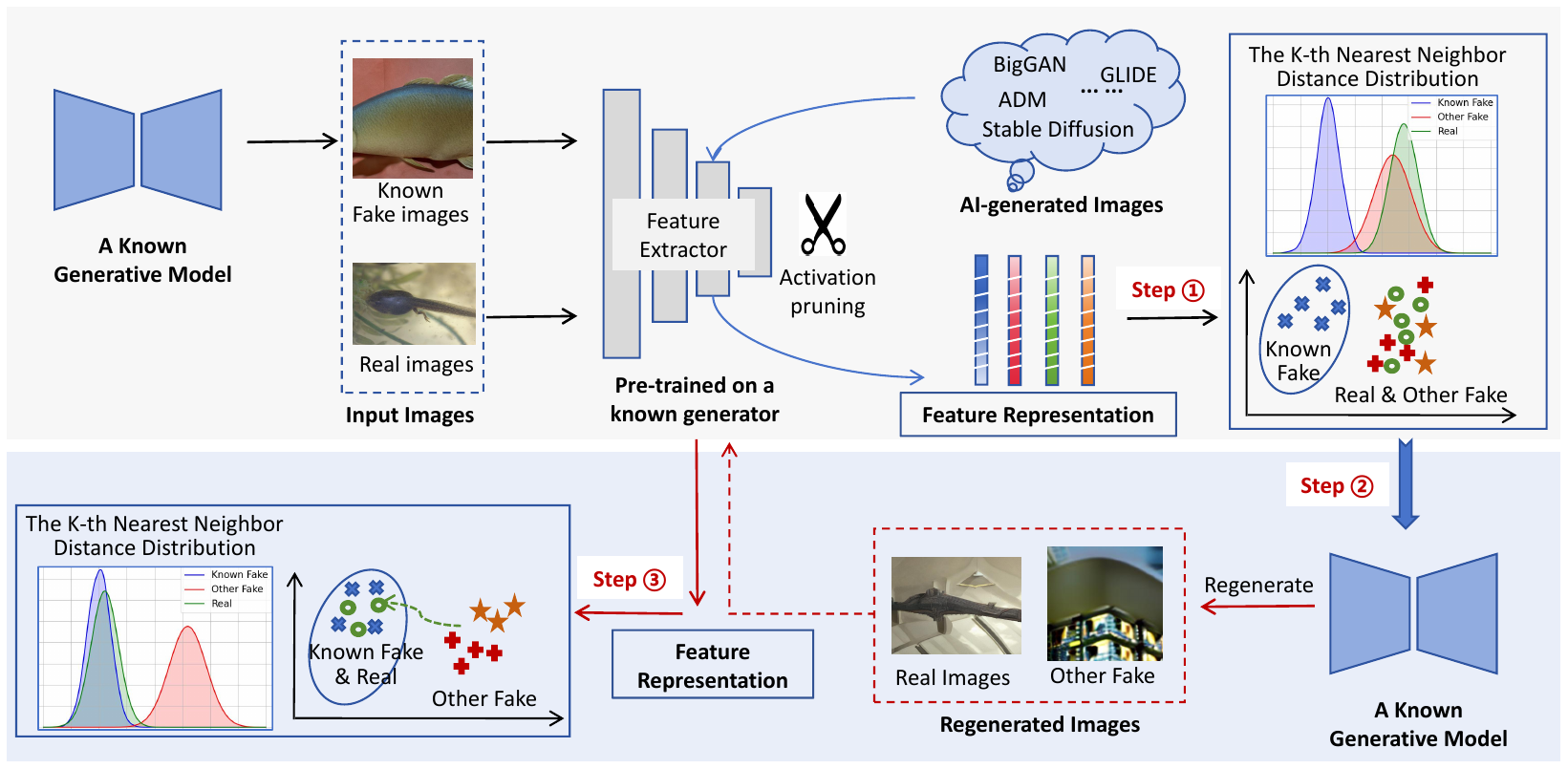}
  \caption{ The overall framework of our PDA. It consists of three key steps:  1) filtering out known fakes by measuring alignment with the known fake distribution in the raw feature space; 2) regenerating the remaining images—a mixture of real and unknown fake samples—using a known generator; and 3) distinguishing real images from unknown fakes in the regenerated feature space based on deep KNN distances and a threshold-based criterion.} 
\label{fig:framework}
\end{figure*}

\subsection{Detection of AI-generated Images}
The detection of AI-generated images has become a critical research topic due to the widespread misuse of generative technologies~\cite{sha2023fake,deng2024towards,cozzolino2024zero,yan2023deepfakebench}. 
Early detection methods primarily focused on identifying generator-specific artifacts, such as pixel-level inconsistencies, using handcrafted features or traditional classifiers~\cite{marra2019gans,li2020face,mccloskey2019detecting}. 
While effective for low-quality or early-generation fakes, these approaches exhibited limited generalization to more advanced generative models. 
With the rise of deep learning, Convolutional Neural Networks (CNNs) have been widely adopted for AI-generated image detection due to their ability to automatically learn discriminative features~\cite{wang2024deepfake,rossler2019faceforensics++}. 
For example, Wang et al.~\cite{wang2020cnn} demonstrate that CNNs trained on ProGAN-generated images can generalize to other GAN-based fakes, benefiting from large-scale training on diverse object categories in LSUN~\cite{yu2015lsun}. 

However, this generalization largely holds within the GAN family and degrades when detectors are applied to images generated by fundamentally different generative paradigms. 
As diffusion models and text-to-image systems have gained prominence, detection has become increasingly challenging~\cite{chen2024drct,stein2024exposing}. 
Unlike GANs, which often leave noticeable artifacts, diffusion models produce highly realistic images with minimal visual discrepancies~\cite{corvi2023intriguing}. 
Zhu et al.~\cite{zhu2024genimage} highlight that classifiers trained exclusively on GAN-based images struggle to generalize to diffusion-generated images, as the two classes exhibit distinct generative fingerprints. 
To mitigate this gap, Wang et al.~\cite{wang2023dire} proposed Diffusion Reconstruction Error (DIRE), exploiting the observation that real images cannot be accurately reconstructed by diffusion models. While effective for diffusion-based generation, DIRE fails to generalize to text-to-image models, as demonstrated in ZeroFake~\cite{sha2024zerofake}. ZeroFake improves upon DIRE by leveraging the differential response of real and fake images to adversarial prompts during inversion and reconstruction, enabling stronger detection performance on text-to-image diffusion models. 
Although ZeroFake does not require retraining, its adversarial prompt optimization and reconstruction pipeline incur substantial computational overhead. 
Moreover, its reliance on case-specific thresholds (e.g., separate settings for fake images and fake artworks) limits its practicality in open-world deployment scenarios.

In summary, existing detection methods face fundamental challenges in generalizing to unknown generative models~\cite{ojha2023towards}. 
As generative models continue to diversify and evolve rapidly, there is a growing need for a universal detection framework that can reliably identify AI-generated images in open-world settings, even when only a single generative model is available during training.

\subsection{Threat Model}
\label{sec:threat_model}

We consider the problem of detecting AI-generated images in an open-world setting, where adversaries may generate images using previously unseen generative models. 
The defender’s objective is to distinguish real images from synthetic ones at test time without retraining the detector as new generators emerge. 
The defender has access to real images and synthetic images generated by a single known generative model, which are used to train a fixed detector. At inference time, the detector observes only the input image and does not know the generative model, training data, or prompts used by the adversary.

Besides, we assume a realistic black-box or gray-box adversary who is aware of the existence of the detection system but does not have access to the detector parameters or the defender’s test-time regeneration process, nor the ability to adaptively query the detector.
Also, the adversary can employ arbitrary generative models, including diffusion-based models and commercial text-to-image systems, and may apply common image post-processing operations such as compression or blurring.

\section{Post-hoc Distribution Alignment}
\label{methodology}

\subsection{Intuition}
\label{Intuition}

As aforementioned, detectors trained on a single fake distribution frequently misclassify unknown fake images as real in open-world settings. 
To overcome this fundamental challenge, we shift the detection focus from modeling diverse and evolving fake distributions to actively transforming the real image distribution. Specifically, we introduce a \textit{regeneration} process (i.e., image-to-image translation using a known generator), which injects consistent and learnable artifacts into regenerated real images, aligning them with the known fake distribution. In contrast, fake images generated by unknown models tend to preserve or amplify their original, incompatible artifact patterns even after regeneration, resulting in persistent misalignment. This discrepancy enables effective separation between real images and unknown fakes. 

From this perspective, we reformulate AI-generated image detection as a \emph{distribution alignment} problem: real images can be aligned to a known fake distribution through regeneration, whereas unknown fake images remain inherently misaligned in feature space. 

\subsection{Framework Overview}
\label{Framework Overview} 
We propose \textit{Post-hoc Distribution Alignment (PDA)}, a model-agnostic detection framework designed to generalize beyond known generative models. 
As illustrated in~\autoref{fig:framework}, PDA first trains a detector using real images and fake images generated by a single known generative model (e.g., Stable Diffusion~\cite{rombach2022high}). 
The detector—excluding its final classification layer—is then repurposed as a feature extractor that captures model-specific artifacts and defines the feature distribution of known fakes. 
PDA decouples the detection process into three stages: (i) early filtering of known fakes in the raw feature space, (ii) regeneration of ambiguous samples via the known generator, and (iii) discrimination between real images and unknown fakes in the regenerated feature space.

This three-stage design allows PDA to adaptively handle inputs based on their position in feature space. 
“Easy” cases, such as known fakes or artifact-similar unknowns, are efficiently filtered in the first stage. 
For “hard” cases where unknown fakes overlap with real images in the raw feature space, regeneration induces alignment for real images but not for fake ones, rendering them separable in the final differentiation stage. 
Overall, PDA is simple, efficient, and broadly applicable, requiring neither semantic priors nor access to the target generator.




\subsection{Feature Extractor Training}
We begin by training a detector \( F_\theta \) to distinguish between real images and fake images generated by a single known generative model \( G(\cdot) \). 
Representative generative models, such as Stable Diffusion V1.4 (SD)~\cite{rombach2022high} from the HuggingFace Hub, are readily accessible. 
The detector is trained on a labeled dataset \(\{(I_i, y_i)\}_{i=1}^N\), where \( y_i \in \{0,1\} \) denotes the ground-truth label (real or fake). The training objective is defined as: 
\begin{equation}
L = \sum_{i=1}^{N} \text{loss}(F_{\theta}(I_i), y_i). 
\end{equation}

After training, the detector—excluding its final classification layer—is used as a feature extractor \( f_{\theta} \), where the penultimate-layer features serve as discriminative embeddings.  
This feature extractor learns to encode model-specific artifacts introduced by the known generator, enabling both threshold-based filtering and subsequent distribution alignment. 
Importantly, PDA does not rely on a specific feature extractor or known generator (see \autoref{backbone} and \autoref{generator}).

\subsection{Reference Set Construction}
To characterize the alignment space, we construct a \emph{reference set} \( \mathbb{Z} \) using feature representations extracted from fake images generated exclusively by a single known generator.  
This reference set captures distributional patterns induced by the known generator’s artifacts, rather than semantic content diversity. 
Given a known fake image \( I \), its feature representation is computed as:
\begin{equation}
\mathbf{x} = f_{\theta}(I),
\label{eq:feature1}
\end{equation}
where \( \mathbf{x} \in \mathbb{R}^d \) denotes the high-dimensional feature embedding. These representations are further refined through activation pruning and dimensionality reduction to enhance discriminative power and suppress irrelevant features.

\mypara{Activation Pruning}
To suppress noisy and spurious activations, we adopt activation pruning~\cite{sun2021react,djurisic2022extremely}, which clips the high-activation dimensions to retain only salient features. Let \( \mathbf{x}_{pruned} \) represent the pruned feature vector, for each feature vector \( \mathbf{x} \), we compute a threshold \( c \) as the 90th percentile of activations and truncate: 
\begin{equation}
\mathbf{x}_{pruned} = \mathcal{P}(\mathbf{x; c}),  
\label{eq:feature2-1}
\end{equation}
\begin{equation}
\mathcal{P}(\mathbf{x} )= \min(\mathbf{h}_{i}(\mathbf{x}), c), 
\label{eq:feature2-2}
\end{equation}
where \( \mathcal{P}(\cdot) \) denotes the pruning operation, and \( \mathbf{h}_{i}(\mathbf{x}) \) represents the activation value of the feature vector \( \mathbf{x} \). The threshold \( c \) is determined as the \( p \)-th percentile of activations for each sample. Following prior work~\cite{sun2021react}, \( p = 90 \) is selected, meaning \( c \) is set so that 90\% of activations lie below the threshold. This approach retains the most informative activations while minimizing noise. 

\mypara{Dimensionality Reduction} 
To facilitates efficient KNN computation while preserving local neighborhood structure, we apply t-SNE~\cite{van2008visualizing} to project pruned features into a low-dimensional space: 
\begin{equation}
\mathbf{z} = \text{t-SNE}(\mathbf{x}_{pruned}), 
\label{eq:feature3}
\end{equation}
where \( \mathbf{z} \in \mathbb{R}^2 \) denotes the reduced feature representation. 
This step improves computational efficiency and interpretability by preserving local geometric relationships. 
Notably, PDA is not tied to any specific dimensionality reduction technique (see~\autoref{ablation_tsne}).


In this way, we construct a \emph{reference set} \( \mathbb{Z} \) that contains feature representations of known fake images (3{,}000 samples): 
\begin{equation}
\mathbb{Z} = \{\mathbf{z}_1, \mathbf{z}_2, \dots, \mathbf{z}_n\}, 
\label{eq:knn1}
\end{equation}
where \( n \) denotes the number of known fake images and \( \mathbf{z}_i \) is the feature representation of the \( i \)-th fake image.
This reference set is used to evaluate the similarity of test images, determining how closely they align with the distribution of known fakes.

\subsection{Threshold Determination}
To distinguish real images from fake ones during inference, we calibrate a detection threshold \( \tau \) using regenerated real images. 
Specifically, a held-out set of real images is passed through the known generator \( G(\cdot) \) to produce pseudo-fake images, which inherit the same artifacts as the training fakes and align closely with the reference distribution. 
These pseudo-fake samples are processed by the feature extractor, activation pruning, and dimensionality reduction, yielding:
\begin{equation}
\mathbb{Z}^{'} = \{ \mathbf{z}_1^{'}, \dots, \mathbf{z}_m^{'} \}. \label{pseudo-fake features}
\end{equation}

To determine an alignment threshold, we compute the \( k \)-nearest neighbor (kNN) distance between each \( \mathbf{z}' \in \mathbb{Z}' \) and the reference set \( \mathbb{Z} \). 
Concretely, for each \( \mathbf{z}' \), we compute the Euclidean distances \( \|\mathbf{z}' - \mathbf{z}_i\|_2 \) for all \( \mathbf{z}_i \in \mathbb{Z} \), and then sort them in ascending order of distance and define the \( k \)-NN distance of \( \mathbf{z}^{'} \) by:
\begin{equation}
d_k(\mathbf{z}^{'}) = ||\mathbf{z}^{'} - \mathbf{z}_k||_2,  
\end{equation}
where \( \mathbf{z}_k \) is the \( k \)-th nearest neighbor of \( \mathbf{z}' \) in \( \mathbb{Z} \).  
Repeating this process for all elements in \( \mathbb{Z}^{'} \), we obtain a set of \( k \)-NN distances. 

Finally, we sort these distances in ascending order and set the threshold $\tau$ at the 95th percentile of the distances, following prior work~\cite{sun2021react,yu2025learning}. This calibration ensures that 95\% of regenerated real images are considered aligned with the known fake distribution, enabling high-confidence separation (see~\autoref{threshold}). 
Importantly, the threshold is calibrated without access to unseen test data and is independent of the generative model used by adversaries, facilitating robust open-world deployment.


\begin{algorithm}[!ht]
\caption{PDA for AI-Generated Image Detection}
\label{alg1}
\renewcommand{\algorithmicrequire}{\textbf{Input:}}
\renewcommand{\algorithmicensure}{\textbf{Output:}}
\begin{algorithmic}[1]
\REQUIRE Test image \( I_{\text{test}} \); known fake images \( \{I_i^{\text{fake}}\}_{i=1}^N \); pre-trained feature extractor \( f_{\theta} \); known generative model \( G(\cdot) \); pruning threshold percentile \( p \); number of nearest neighbors \( k \)
\ENSURE Predicted label \( \hat{y} \in \{\text{Real}, \text{Fake}\} \)

\vspace{4pt}
\STATE \textbf{Reference Set Construction:}
\FOR{each \( I_i^{\text{fake}} \)}
    \STATE Extract feature: \( \mathbf{x}_i = f_{\theta}(I_i^{\text{fake}}) \)
    \STATE Activation pruning: \( \mathbf{x}_i^{\text{pruned}} = \mathcal{P}(\mathbf{x}_i; p) \)
    \STATE Dimensionality reduction: \( \mathbf{z}_i = \text{t-SNE}(\mathbf{x}_i^{\text{pruned}}) \)
\ENDFOR
\STATE Reference feature set: \( \mathbb{Z} = \{\mathbf{z}_1, \dots, \mathbf{z}_N\} \)

\vspace{4pt}
\STATE \textbf{Threshold Determination:}
\STATE Regenerate real images: \( \{I_j^{\text{real}}\}_{j=1}^M \rightarrow \{I_j^{\text{pseudo}} = G(I_j^{\text{real}})\} \)
\FOR{each \( I_j^{\text{pseudo}} \)}
    \STATE Compute \( \mathbf{z}_j' \) via same steps as above
    \STATE Compute \( k \)-NN distance: \( d_k(\mathbf{z}') = ||\mathbf{z}' - \mathbf{z}_k||_2 \)
\ENDFOR
\STATE Set threshold \( \tau \) as 95th percentile of \( \{d_k(\mathbf{z}_j')\}_{j=1}^M \)

\vspace{4pt}
\STATE \textbf{Detection for Test Image \( I_{\text{test}} \):}
\STATE Extract and reduce \( \mathbf{z}^* \) using same pipeline
\STATE Compute \( k \)-NN distance: \( d_k(\mathbf{z}^*) = ||\mathbf{z}^* - \mathbf{z}_k||_2 \)
\IF{\( d_k(\mathbf{z}^*) \leq \tau \)}
    \STATE \textbf{return} Fake
\ELSE
    \STATE Generate pseudo-fake: \( I_{\text{pseudo}} = G(I_{\text{test}}) \)
    \STATE Extract and reduce: \( {z}^{*}_{pseudo} \)
    
    \STATE Compute \( k \)-NN distance: \( d_k({z}^{*}_{pseudo}) = ||{z}^{*}_{pseudo} - \mathbf{z}_k||_2 \)
    \IF{\( d_k({z}^{*}_{pseudo}) \leq \tau \)}
        \STATE \textbf{return} Real
    \ELSE
        \STATE \textbf{return} Fake
    \ENDIF
\ENDIF
\end{algorithmic}
\end{algorithm}

\subsection{Inference-time Detection} 
\label{three-step detection}
This section describes the inference-time detection of PDA and how the proposed three-stage pipeline enables reliable detection of AI-generated images from previously unknown generators.


\textbf{Step \filledcircled[\small]{1} – Raw-space Filtering.} 
Given a test image \( \mathbf{I}_{test} \), we first extract its feature representation \( \mathbf{z}^* \) using the feature extractor \( f_\theta \), and compute its \( k \)-nearest neighbor distance \( d_k(\mathbf{z}^*) \) to the reference set \( \mathbb{Z} \). 
If \( d_k(\mathbf{z}^*) \leq \tau \), the image is classified as \textbf{fake}, since it aligns with the feature-space distribution of known fake images.

\textbf{Step \filledcircled[\small]{2} – Regeneration.} 
Images that are not filtered in the first stage—corresponding to a mixture of real and unknown fake images—are passed through the known generative model \( G(\cdot) \) to obtain regenerated pseudo-fake samples $\mathbf{x}_{\text{pseudo}} = f_{\theta}(G(\mathbf{I}_{\text{test}}))$. 

\textbf{Step \filledcircled[\small]{3} – Differentiation.}
We then repeat the feature extraction process and compute the \( k \)-NN distance to the reference set: 
\begin{equation}
\begin{aligned}
    &\mathbf{x}_{pseudo\_pruned} = \mathcal{P}(\mathbf{x}_{pseudo}; c), \\
    & \mathbf{z}^{*}_{pseudo} = \text{t-SNE}(\mathbf{x}_{pseudo\_pruned}), \\
    &d_k(\mathbf{z}^{*}_{pseudo})  = ||\mathbf{z}^{*}_{pseudo} - \mathbf{z}_k||_2. 
\end{aligned}
\label{eq:pseudo_all}
\end{equation}
The \( d_k(\mathbf{z}^{*}_{pseudo}) \) is subsequently compared with the threshold \( \tau \). If the distance is smaller than \( \tau \), the image is classified as \textbf{real}, as only real images produce pseudo-fake samples with the same artifacts and features as known fake images. Conversely, if the distance is larger than \( \tau \), the image is classified as an unknown \textbf{fake}.

The three-step detection procedure is formalized as follows:
\begin{equation}
\hat{y} =
\begin{cases}
\textbf{Fake}, & d_k(\mathbf{z}^{*})\leq \tau \\
\textbf{Real}, & d_k(\mathbf{z}^{*}) > \tau \text{ and }  d_k(\mathbf{z}^{*}_{pseudo})  \leq \tau \\
\textbf{Fake}, & d_k(\mathbf{z}^{*}) > \tau \text{ and }  d_k(\mathbf{z}^{*}_{pseudo})   > \tau
\end{cases}
\label{eq:knn3}
\end{equation}

This three-step strategy allows PDA to adaptively classify known fakes, unknown fakes, and real samples without retraining or fine-tuning on new generative models. The detailed procedure of PDA is presented in Algorithm \ref{alg1}.

\begin{tcolorbox}[title=Remark,colback=gray!5!white,colframe=gray!75!black,fonttitle=\bfseries]
PDA reformulates detection as a distribution alignment task by regenerating test images through a known generative model. 
Reals become aligned through regeneration, while unknown fakes — preserving conflicting or mixed artifacts — remain misaligned. This enables generalized detection even under diverse, open-world generative threats.
\end{tcolorbox}

\section{Experiments}
\subsection{Experimental Setup}
\begin{table*}[!t]
\centering
\caption{Overview of the 16 generative models used in our experiments.}
\label{tab:generative_models_detailed}
\scalebox{0.8}{  
\begin{tabular}{llllll}
\toprule
\textbf{Model Name} & \textbf{Abbreviation} & \textbf{Architecture Type} & \textbf{Typical Resolution} & \textbf{Source Dataset}  \\ \midrule
ProGAN~\cite{karras2017progressive} & ProGAN & GAN (Image-to-Image)& 256$\times$256 & LSUN  \\ 
StyleGAN~\cite{karras2019style} & StyleGAN & GAN (Image-to-Image)& 256$\times$256 & LSUN \\
BigGAN~\cite{brock2018large} & BigGAN & GAN (Image-to-Image) & 256$\times$256 (up to 512$\times$512) & ImageNet  \\
CycleGAN~\cite{zhu2017unpaired} & CycleGAN & GAN (Image-to-Image) & 256$\times$256 & ImageNet \\
StarGAN~\cite{choi2018stargan} & StarGAN & GAN (Image-to-Image) & 256$\times$256 & CelebA  \\ 
GauGAN~\cite{park2019semantic} & GauGAN & GAN (Image-to-Image) & 256$\times$256 & COCO \\
StyleGAN2~\cite{karras2020analyzing} & StyleGAN2 & GAN (Image-to-Image) & 256$\times$256 (up to 1024$\times$1024) & LSUN  \\ 
WhichFaceIsReal~\cite{whichfaceisreal_dataset} (StyleGAN-based) & WFIR & GAN (Face Generation Focus) & 1024$\times$1024 & FFHQ \\ 
Midjourney~\cite{midjourney_website} (Commercial API)  & Midjourney & Diffusion-based (Text-to-Image) & $\geq$1024$\times$1024 & ImageNet   \\ 
DALL·E 2~\cite{ramesh2022hierarchical} (Commercial API) & DALL·E 2 & Diffusion-based (Text-to-Image) & 256$\times$256 (up to 1024$\times$1024) &ImageNet   \\ 
SDXL~\cite{podell2023sdxl} & SDXL & Latent Diffusion (Text-to-Image) & 1024$\times$1024 & COCO \\ 
Stable Diffusion v1.4~\cite{rombach2022high} & SD & Latent Diffusion (Text-to-Image) & 512$\times$512 & ImageNet  \\
GLIDE~\cite{nichol2021glide} & GLIDE & Diffusion-based (Text-to-Image) & 256$\times$256 &  ImageNet  \\
Vector Quantized Diffusion Model~\cite{gu2022vector} & VQDM & Diffusion + VQ (Text-to-Image) & 256$\times$256 &  ImageNet  \\
Ablated Diffusion Model~\cite{dhariwal2021diffusion} & ADM & Diffusion-based (Text-to-Image) & 256$\times$256 & ImageNet  \\
Wukong~\cite{wukong2022} & Wukong & Diffusion-based (Chinese Text-to-Image) & 512$\times$512 & ImageNet \\
\bottomrule
\end{tabular}%
}
\end{table*}

\subsubsection{Datasets and Generation Models}
To rigorously evaluate the generalization capabilities and robustness of our PDA method, the experiments are conducted on two comprehensive datasets: 

\begin{itemize}
    \item \textbf{GenImage} dataset~\cite{zhu2024genimage} is a large-scale collection specifically curated for evaluating the detection of AI-generated images, comprising over 2.68 million images in total. This includes 1.33 million real images from the widely recognized ImageNet dataset~\cite{deng2009imagenet}, ensuring a diverse representation of real-world visual content across 1,000 categories. The remaining 1.35 million images are synthetic, generated by six distinct generative models spanning a range of architectures—including GANs, diffusion-based models, and text-to-image systems. 

    \item \textbf{AIGCDetect} dataset~\cite{zhong2023patchcraft} provides a diverse set of synthetic images from numerous contemporary generative models. It aggregates approximately 151,500 images generated by 17 different models spanning various architectures, and incorporates outputs from commercial text-to-image services such as Midjourney~\cite{midjourney_website}. The images exhibit varied resolutions and are synthesized based on diverse source datasets such as LSUN~\cite{yu2015lsun}, ImageNet~\cite{deng2009imagenet}, and COCO~\cite{lin2014microsoft}. AIGCDetec dataset incorporates several models not present in GenImage, such as StyleGAN2~\cite{karras2020analyzing} and SDXL~\cite{podell2023sdxl}, rendering it crucial for comprehensively evaluating PDA's adaptability to the rapidly evolving landscape of AI image generation.  
\end{itemize} 

In total, our evaluation involves \textbf{16 distinct generative models}, comprising 8 GAN-based and 8 diffusion-based models, including commercial text-to-image APIs. Detailed model information is summarized in~\autoref{tab:generative_models_detailed}, with additional descriptions provided in~\autoref{appendix_models}. Following the open-world detection protocol, \textbf{only one model} is treated as known generator and used to pretrain the feature extractor, while the remaining 15 models are strictly unseen during training, enabling a rigorous evaluation of PDA’s robustness and generalization to previously unknown generators.  

\begin{table*}[!t]
  \centering
  \caption{Detection accuracy of PDA across 16 diverse generative models}
  \label{tab:effectiveness} 
  \scalebox{0.83}{  
  \begin{tabular}{l r r r r r r r }
    \toprule
    \textbf{Fake Data Source}& \textbf{Benchmark} & \textbf{CNNDetection~\cite{wang2020cnn}} & \textbf{DIRE}~\cite{wang2023dire} & \textbf{Ojha et al.}~\cite{ojha2023towards} & \textbf{ZeroFake}~\cite{sha2024zerofake} & \textbf{DRCT}~\cite{chen2024drct} & \textbf{PDA (Ours)} \\
    \midrule
    \cmidrule(r){1-1} 
    ProGAN        & 67.25\%         & 99.70\%         & 50.05\%         & 93.85\%         & 71.00\%         & 89.45\%         & \textbf{98.70}\% \\
    StyleGAN      & 64.20\%         & 68.75\%         & 58.10\%         & 84.00\%         & 65.10\%         & 87.00\%         & \textbf{96.13}\% \\
    BigGAN        & 63.42\%         & 73.70\%         & 59.72\%         & 89.60\%         & 82.55\%         & 87.70\%         & \textbf{98.19}\% \\
    CycleGAN      & 70.30\%         & 84.55\%         & 46.20\%         & 92.90\%         & 66.35\%         & 89.25\%         & \textbf{98.14}\% \\
    StarGAN       & 93.75\%         & 84.90\%         & 62.90\%         & 93.00\%         & 78.10\%         & 91.05\%         & \textbf{98.14}\% \\
    GauGAN        & 51.70\%         & 82.65\%         & 49.25\%         & 93.55\%         & 80.89\%         & 77.10\%         & \textbf{98.34}\% \\
    StyleGAN2     & 54.20\%         & 67.60\%         & 54.15\%         & 73.20\%         & 67.35\%         & 85.00\%         & \textbf{96.59}\% \\
    WFIR          & 52.30\%         & 56.35\%         & 62.45\%         & 87.20\%         & 53.80\%         & 90.05\%         & \textbf{97.79}\% \\
    \rowcolor{aprowcolor}
    \textbf{Avg. (GAN-based models)} & 64.64\% & 77.28\% & 55.35\% & 88.41\% & 70.64\% & 87.08\% & \textbf{97.75\%} \\
    \midrule
    \cmidrule(r){1-1} 
    Midjourney    & 52.90\%         & 50.95\%         & 56.30\%         & 55.90\%         & 66.52\%         & 86.30\%         & \textbf{88.99}\% \\
    DALL·E2       & 65.30\%         & 49.75\%         & 57.95\%         & 50.15\%         & 82.50\%         & 86.00\%         & \textbf{97.99}\% \\
    SDXL          & 52.30\%         & 50.40\%         & 55.10\%         & 59.35\%         & 66.46\%         & 76.50\%         & \textbf{95.87}\% \\
    SD            & \textbf{98.53}\% & 50.15\%         & 51.22\%         & 63.90\%         & 85.63\%         & 90.80\%         & 96.00\% \\
    GLIDE         & 64.27\%         & 52.62\%         & 60.18\%         & 62.70\%         & 85.00\%         & 86.20\%         & \textbf{98.09}\% \\
    VQDM          & 59.00\%         & 52.17\%         & 52.32\%         & 85.60\%         & 69.38\%         & 87.40\%         & \textbf{97.87}\% \\
    ADM           & 54.70\%         & 51.17\%         & 54.97\%         & 67.00\%         & 83.00\%         & 75.40\%         & \textbf{98.12}\% \\
    Wukong        & \textbf{94.65}\% & 50.15\%         & 52.75\%         & 71.10\%         & 78.37\%         & 90.50\%         & 92.10\% \\
    \rowcolor{aprowcolor}
    \textbf{Avg. (Diffusion-based models)} & 67.71\% & 50.92\% & 55.10\% & 64.46\% & 77.11\% & 84.89\% & \textbf{95.63\%} \\
    \midrule
    \rowcolor{aprowcolor}
    \textbf{AP (Overall)}   & 66.18\%   & 64.10\%   & 55.23\%   & 76.44\%   & 73.88\%   & 85.98\%   & \textbf{96.69\%} \\
    \bottomrule
  \end{tabular}%
  } 
\end{table*}

\subsubsection{Evaluation Metrics}
Following existing generated-image detection methods~\cite{sha2023fake,wang2023dire,ojha2023towards,tan2024rethinking}, we report \textbf{accuracy (ACC)} and \textbf{average precision (AP)} to evaluate the detectors. 
Given that our method performs two separate threshold-based classification—one in the raw feature space and another in the regenerated feature space—the overall ACC is defined as follows:
\begin{equation}
\text{ACC} \left ( \% \right ) = 100\times  \frac{N_{\text{correct1}} + N_{\text{correct2}}}{N_{\text{total}}},
\label{eq:acc}
\end{equation}
where \( N_{\text{correct1}} \) denotes the number of samples correctly classified during the initial filtering step, and \( N_{\text{correct2}} \) refers to the number of correct classifications based on the regenerated features. \( N_{\text{total}} \) is the total number of test samples. Note that all evaluations use \textbf{balanced sets} of real and fake images.

In addition to quantitative metrics, we provide \textbf{qualitative visualizations} of the feature space using t-SNE and KNN distance distributions. The results reveal how well our method separates real and fake images, providing insights into the effectiveness of the distribution alignment process.

\subsubsection{Baseline methods}
We compare PDA with several state-of-the-art
methods: 
1) \textbf{Benchmark} (pre-trained feature extractor~\cite{he2016deep}) is trained on fake images generated by SD and real images. This baseline serves as a reference to illustrate the generalization capability of a detector trained on a single generative model. It highlights the limitations of traditional approaches when faced with unknown fake images generated by different models. 
2) \textbf{CNNDetection~\cite{wang2020cnn}} is trained on ProGAN-generated images with simple data augmentation techniques. The classifier has been shown to generalize well to other GAN-based generated images.
3) \textbf{DIRE~\cite{wang2023dire}} distinguishes real and fake images by using the reconstruction error from a pre-trained diffusion model. It leverages the observation that fake images generated by diffusion models have smaller reconstruction errors compared to real images.
4) \textbf{Ojha et al.~\cite{ojha2023towards}} propose to perform real-vs-fake classification without explicit training for this task. Instead, they utilize the feature space of large pre-trained vision-language models\cite{dosovitskiy2020image}, demonstrating that simple nearest neighbor classification achieves strong generalization in detecting fakes from diverse, unknown generators.
5) \textbf{ZeroFake~\cite{sha2024zerofake}} uses a perturbation-based DDIM inversion technique with prompt guidance to distinguish real from fake images, achieving significantly better performance than DIRE.
6) \textbf{DRCT~\cite{chen2024drct}} aims to enhance detector generalizability by focusing on hard-to-classify samples. It generates challenging fake images through high-quality diffusion reconstruction and employs contrastive training to guide the model in learning discriminative diffusion artifacts, thereby improving performance on images from unknown models.

\subsection{Detection Performance}
\label{Performance_fake}
We evaluate on balanced test sets with 3,000 real images from ImageNet and 3,000 synthetic images per generator. The detector’s feature extractor is trained on SD-generated images, treating SD as the \textit{known fake}, while images from the remaining 15 generators are used as \textit{unknown fakes} to evaluate cross-model generalization (more implementation details are provided in Appendix~\ref{appendix_implement}).

\subsubsection{Comparison to Baseline Methods}
\autoref{tab:effectiveness} presents the detection accuracy of PDA and six state-of-the-art detection methods across 16 diverse generative models. Overall, PDA consistently achieves the highest ACC under the open-world setting, with an AP of \textbf{96.69\%}, significantly surpassing the strongest baseline (DRCT: \textbf{85.98\%}) by a margin of \textbf{+10.71\%}. This substantial improvement highlights the generalization of PDA without relying on retraining, prompt engineering, or access to unknown models.

The results reveal that existing detection methods tend to suffer from strong dependency on the specific characteristics of their training generators. For instance, CNNDetection performs well on ProGAN (99.70\%), indicating it likely learned ProGAN-specific artifacts during training, but struggles on more modern architectures like SDXL (50.40\%) and Wukong (50.15\%). Similarly, DIRE, which detects outliers through image reconstruction error, exhibits limited generalizability when applied to unknown generators, achieving only 55.23\% overall. Even ZeroFake, a prompt-aware method tailored to diffusion models, shows performance degradation under distribution shift. For example, on VQDM and ADM, PDA outperforms ZeroFake by over 28\% and 15\%, respectively. Moreover, ZeroFake requires computationally expensive inversion and model-specific thresholds, limiting its practicality in latency-sensitive scenario. 

Interestingly, we observe a consistent trend across all detectors: generative models based on GANs are generally easier to detect than those based on diffusion or text-to-image systems. As shown in \autoref{tab:effectiveness}, the average detection accuracy of all methods is higher on GAN-based generators than on diffusion-based ones. For example, CNNDetection achieves an AP of 77.28\% on GANs but drops to 50.92\% on diffusion models. Similarly, DRCT and Ojha et al. also demonstrate better performance on GANs. This phenomenon likely stems from the fact that GAN-generated images tend to exhibit more localized, spatially structured artifacts, such as checkerboard textures or unnatural edges, which are easier for detectors to capture. In contrast, diffusion models generate images through iterative denoising and often produce globally coherent yet subtly perturbed outputs, making detection inherently more challenging.

Despite this, PDA maintains strong and consistent performance across both categories. On GAN-based models, it achieves an average AP of \textbf{97.75\%}, while on diffusion-based models, the performance remains high at \textbf{95.63\%}. This robustness confirms that PDA does not rely on any generator-specific assumptions or handcrafted features, and instead leverages a general post-hoc distribution alignment principle that adapts to diverse generative sources under open-world settings. 

\begin{figure*}[htbp]
    \centering
        \includegraphics[width=0.23\linewidth]{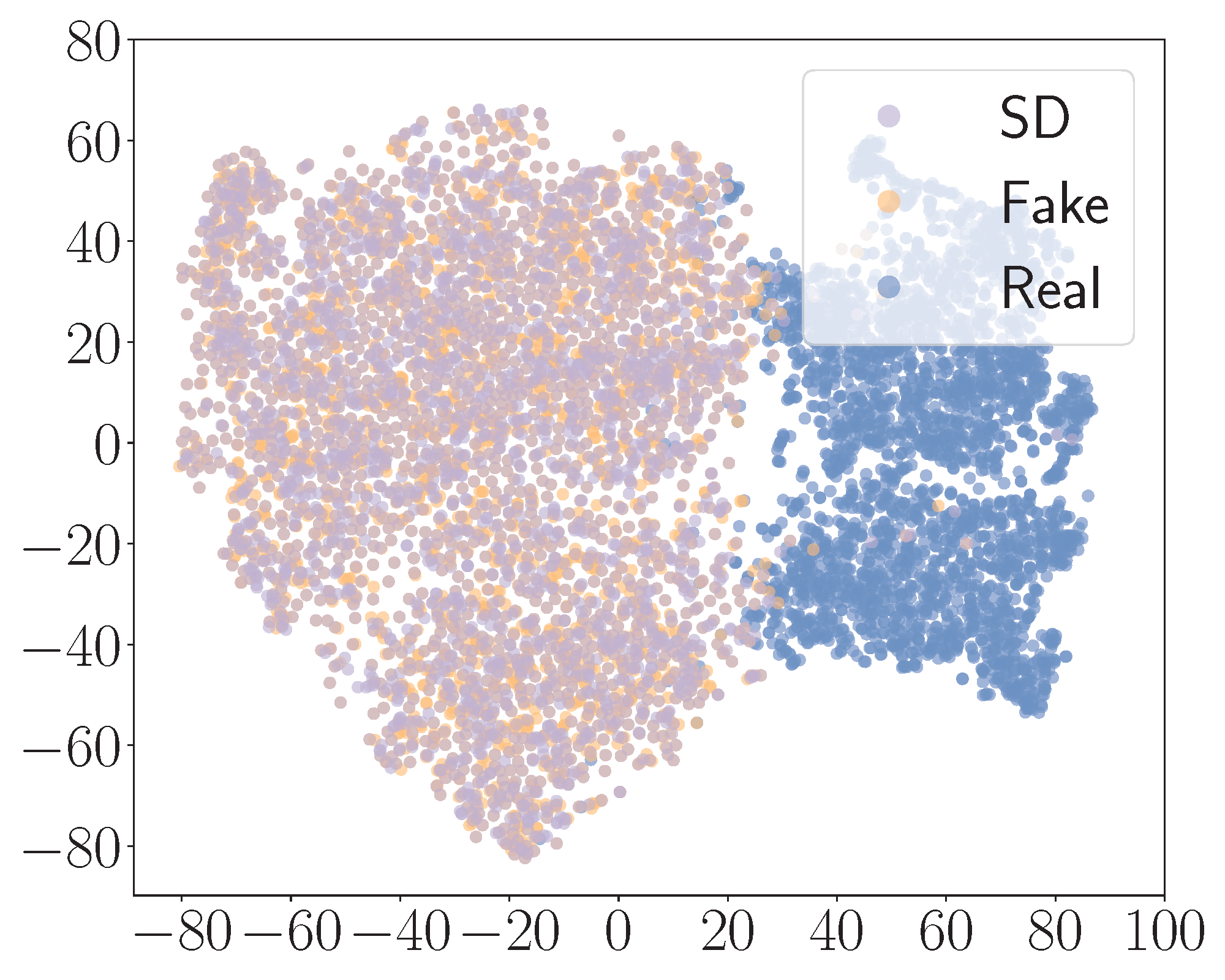}
        \includegraphics[width=0.23\linewidth]{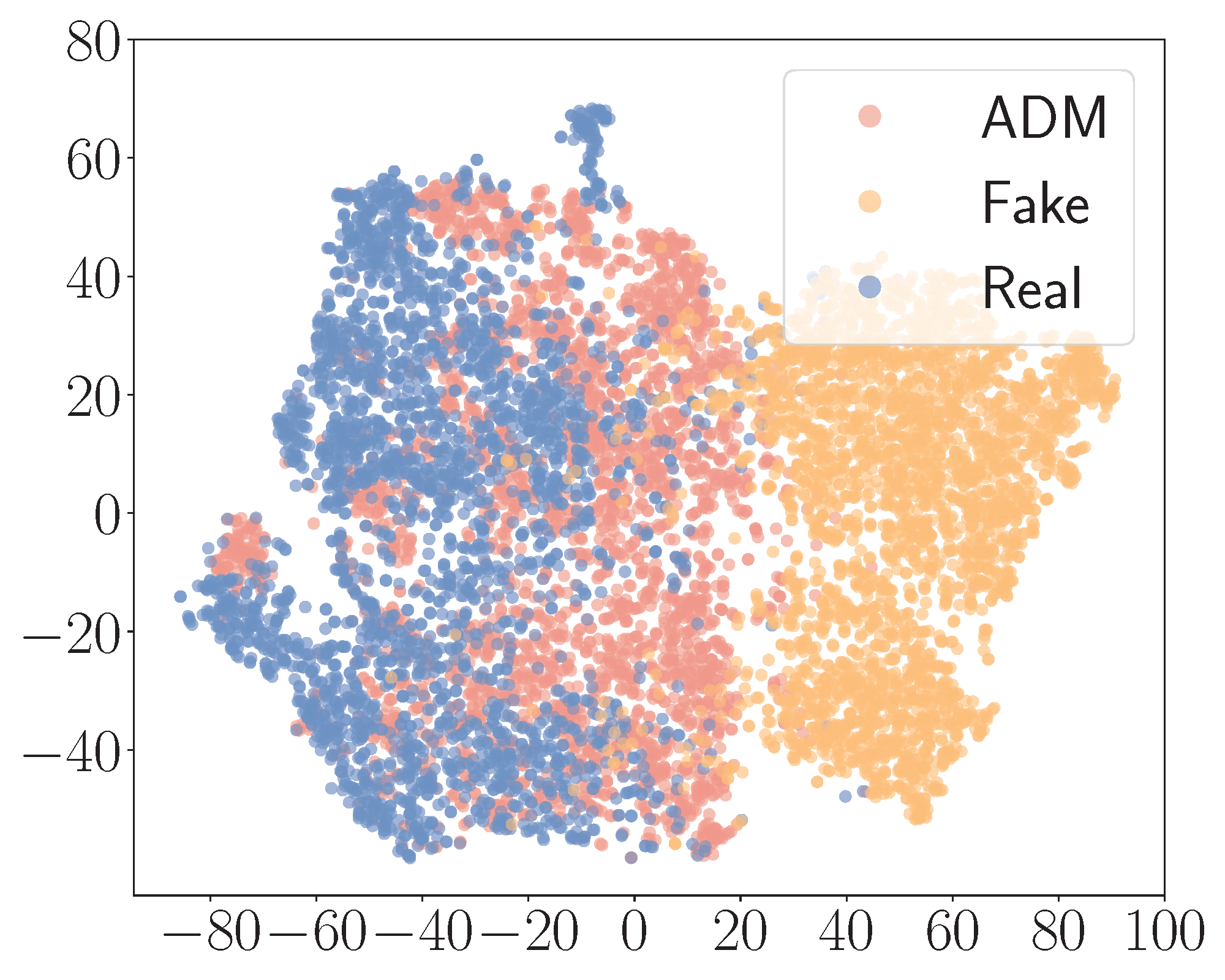}
        \includegraphics[width=0.23\linewidth]{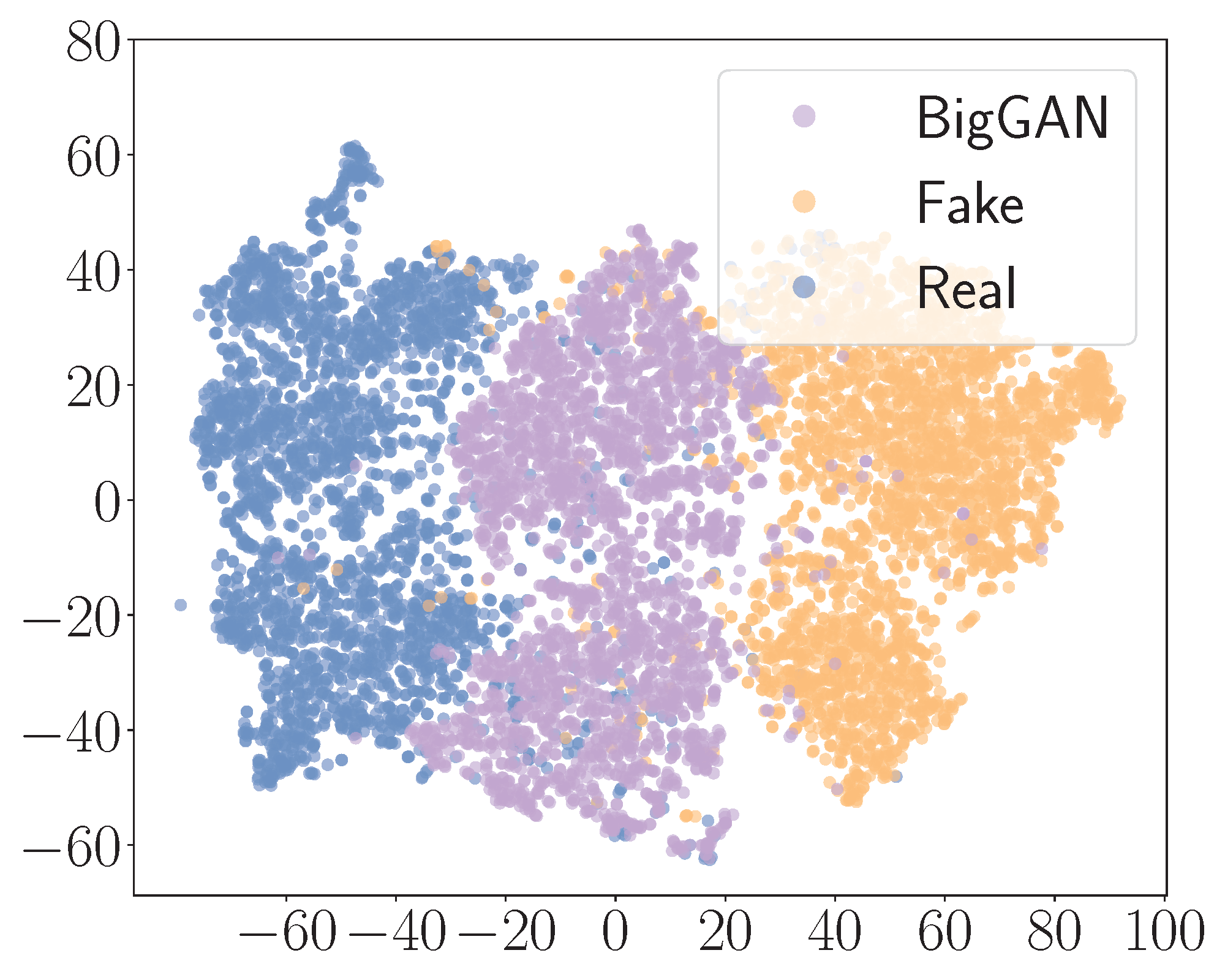}
        \includegraphics[width=0.22\linewidth]{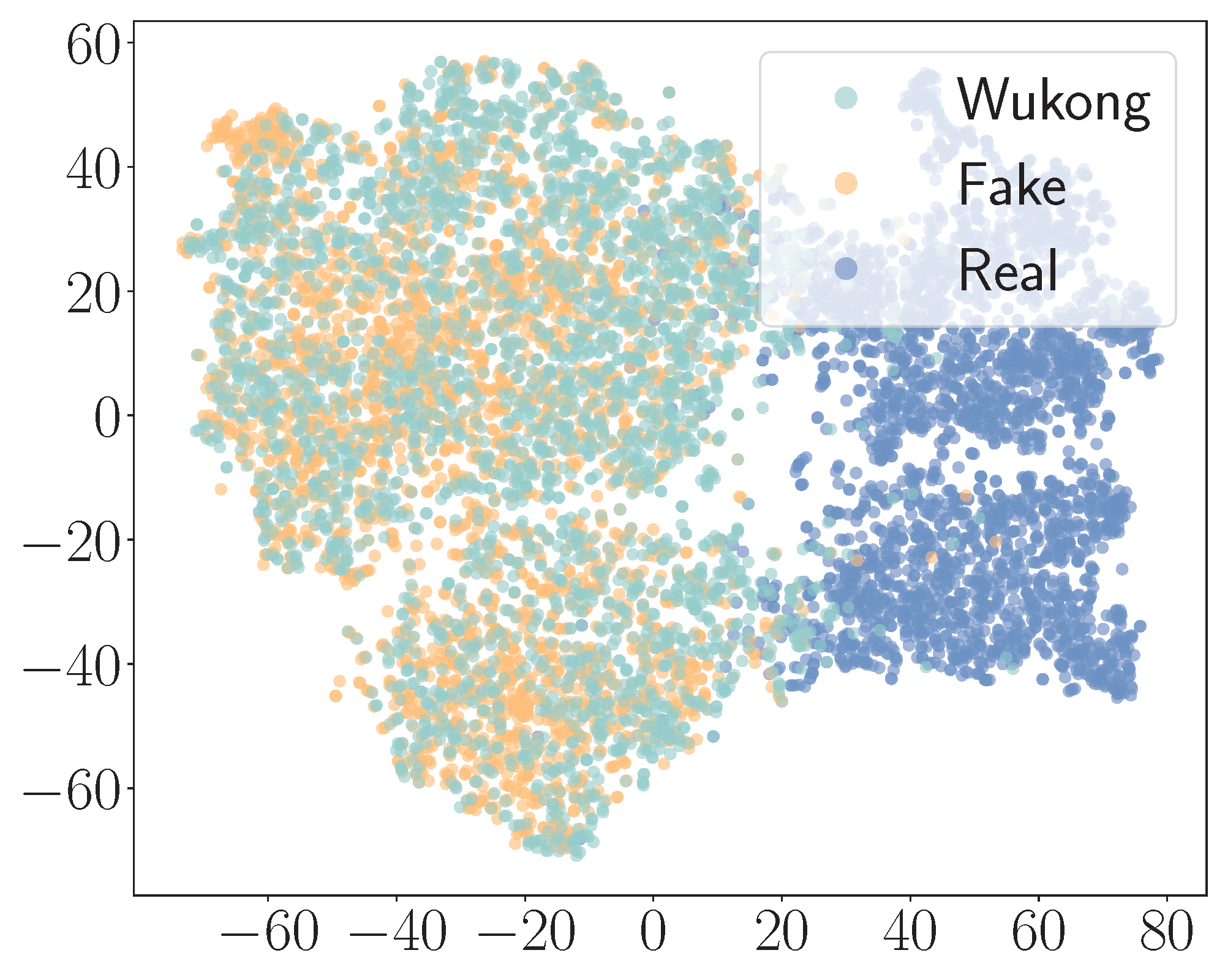}
  \\
     \includegraphics[width=0.23\linewidth]{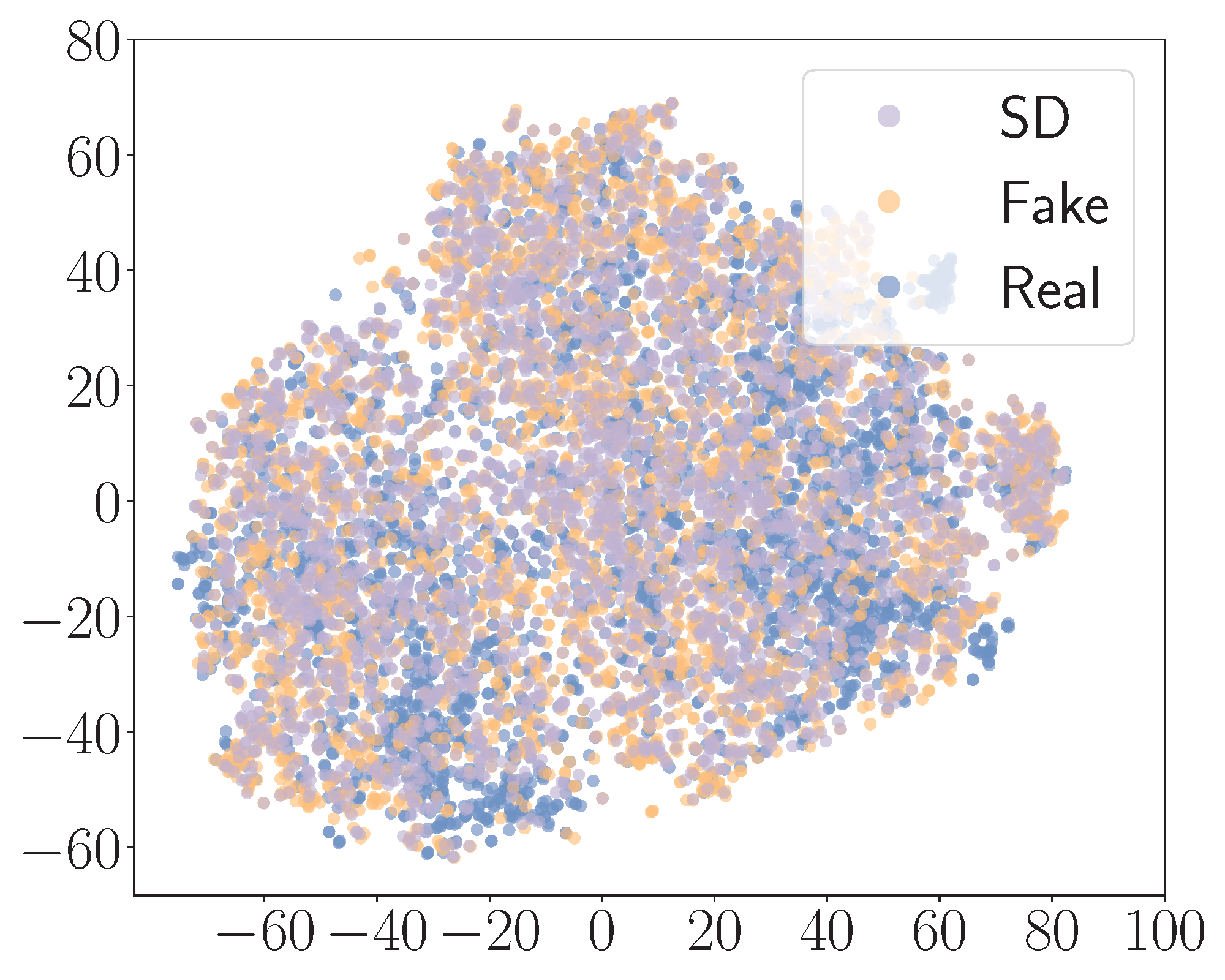}
     \includegraphics[width=0.23\linewidth]{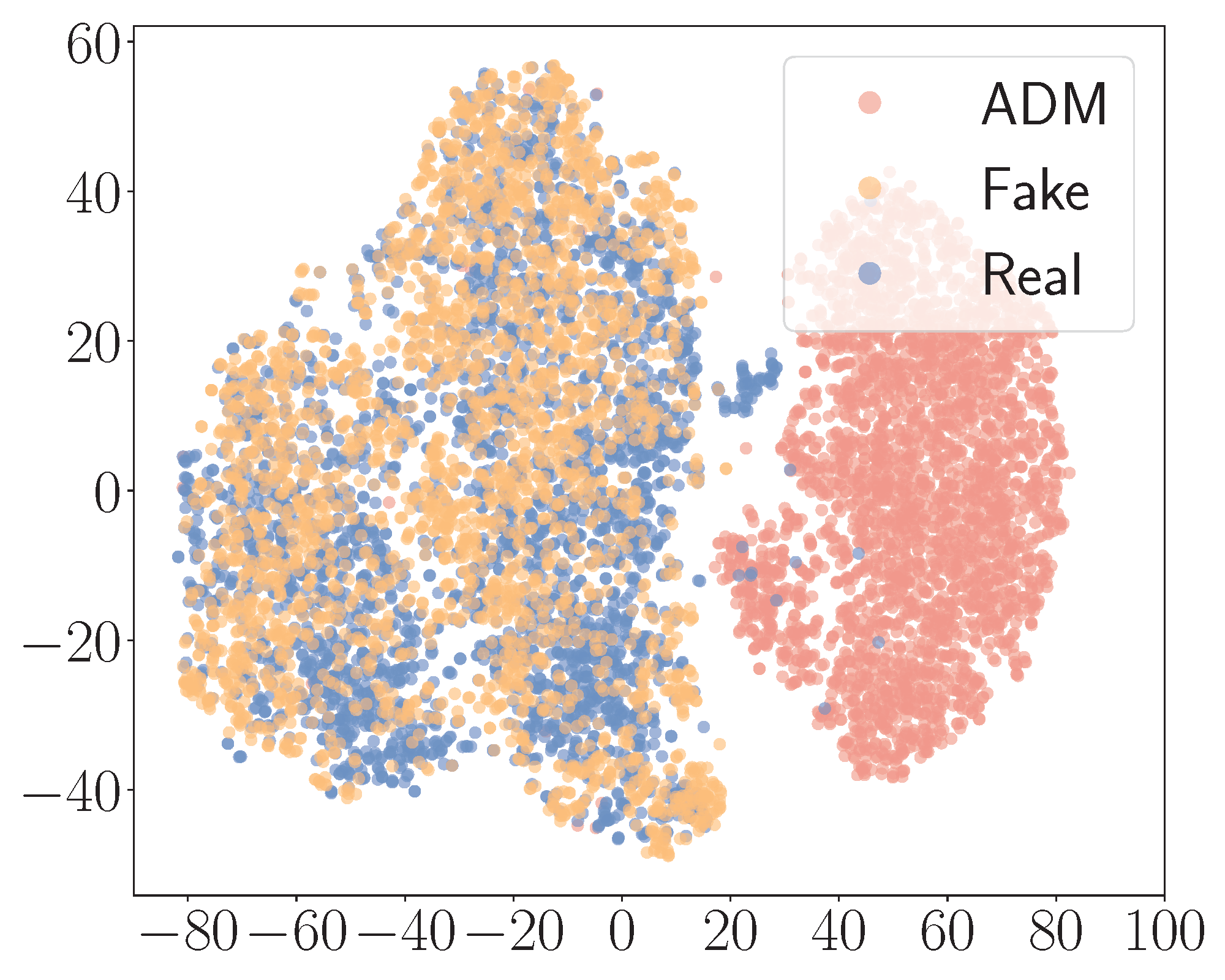} 
     \includegraphics[width=0.23\linewidth]{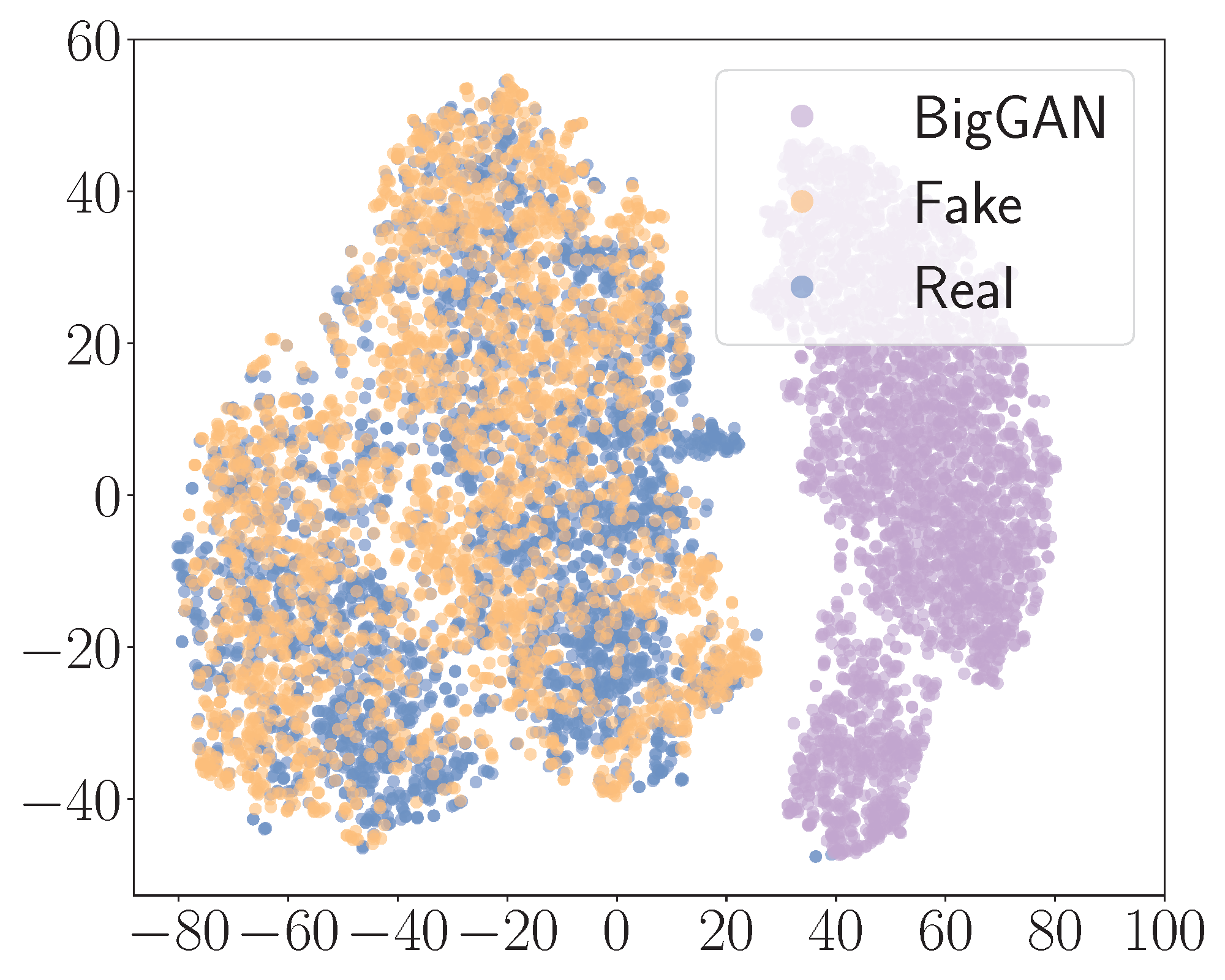}  
     \includegraphics[width=0.23\linewidth]{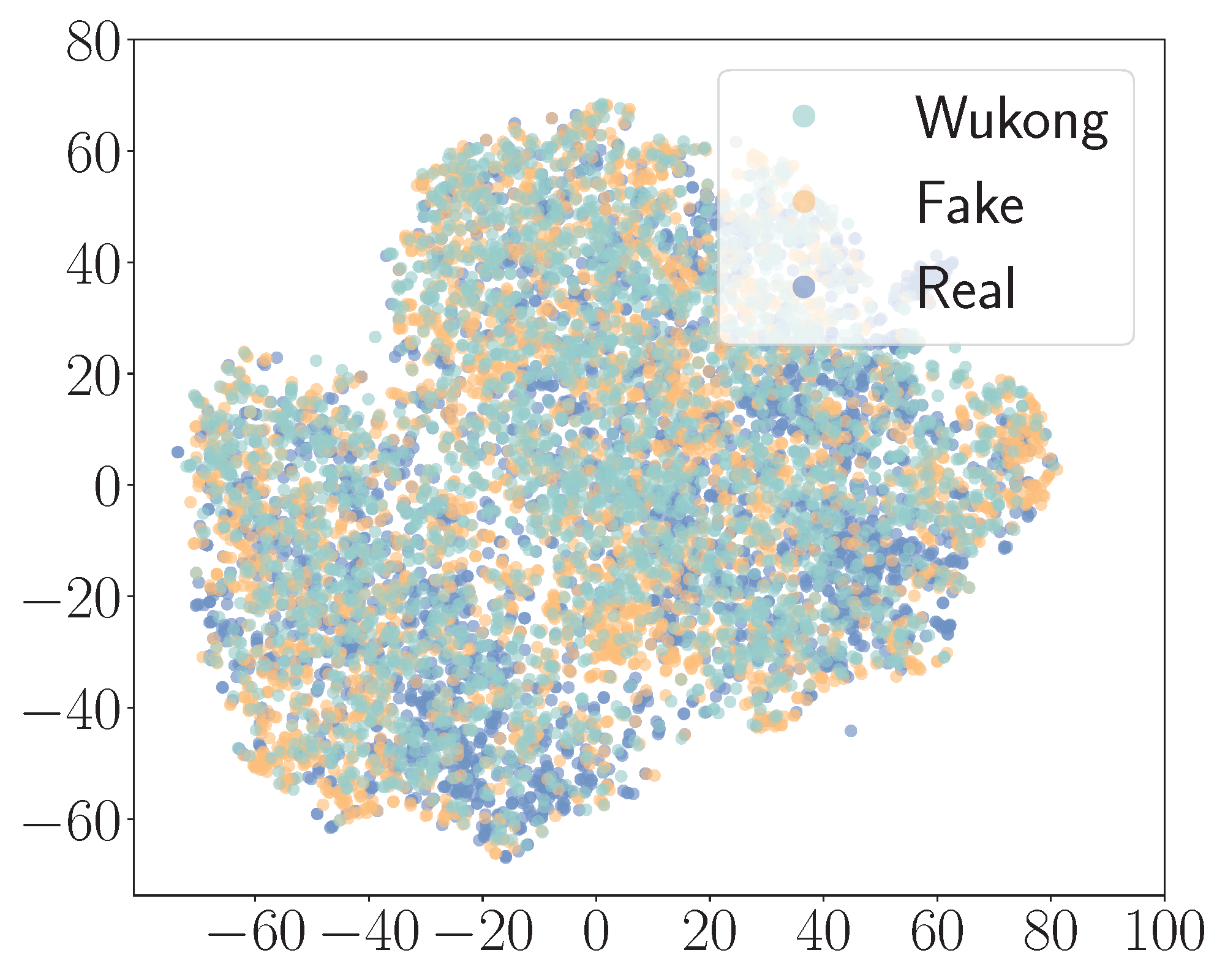}
     \\
\caption{T-SNE visualization. Rows correspond to raw and regenerated feature spaces (“Fake” denotes known fake distribution).}
\label{fig:T-SNE visualization}
\end{figure*}

\begin{figure*}[htbp]
    \centering
    \includegraphics[width=0.23\linewidth]{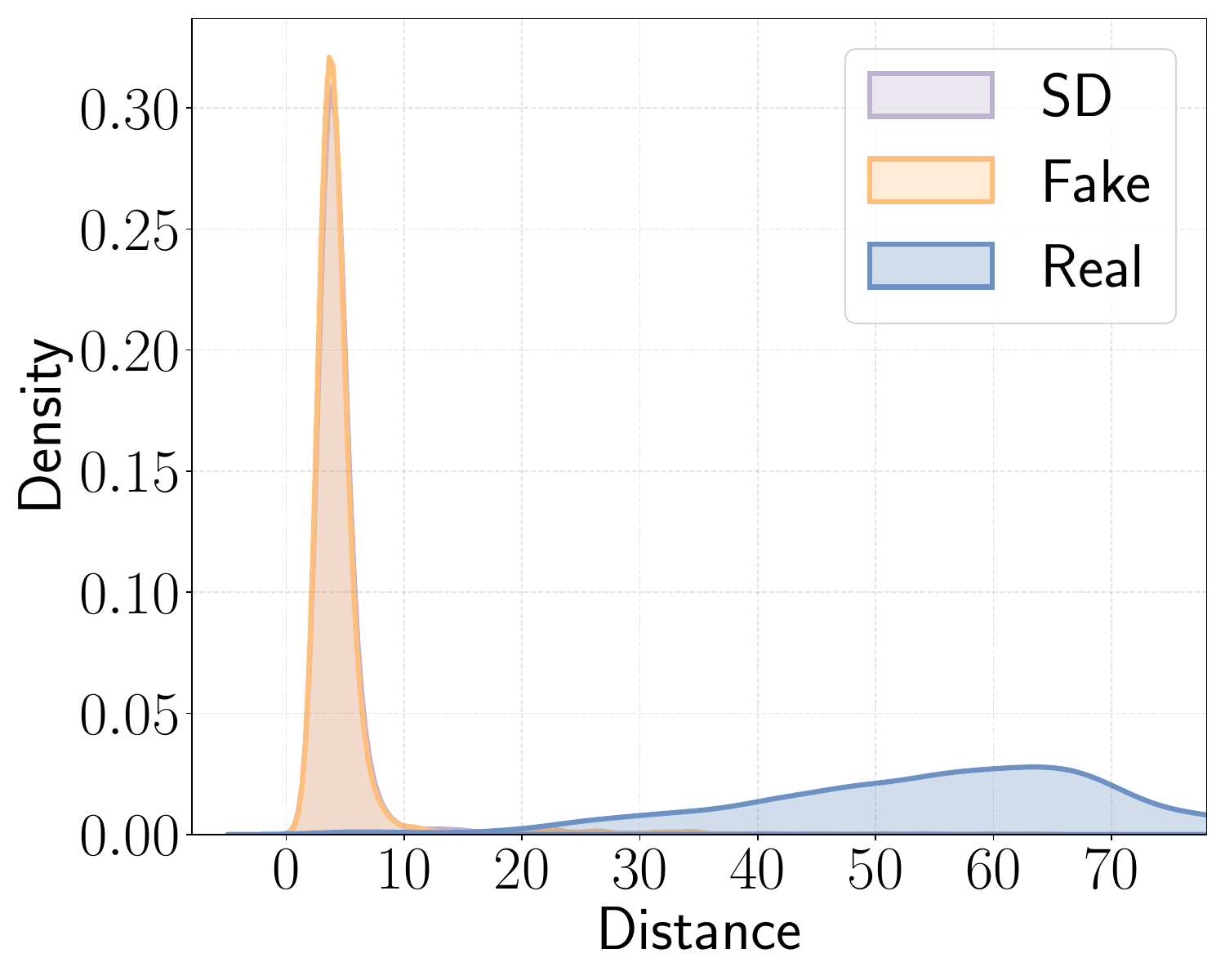} 
    \includegraphics[width=0.23\linewidth]{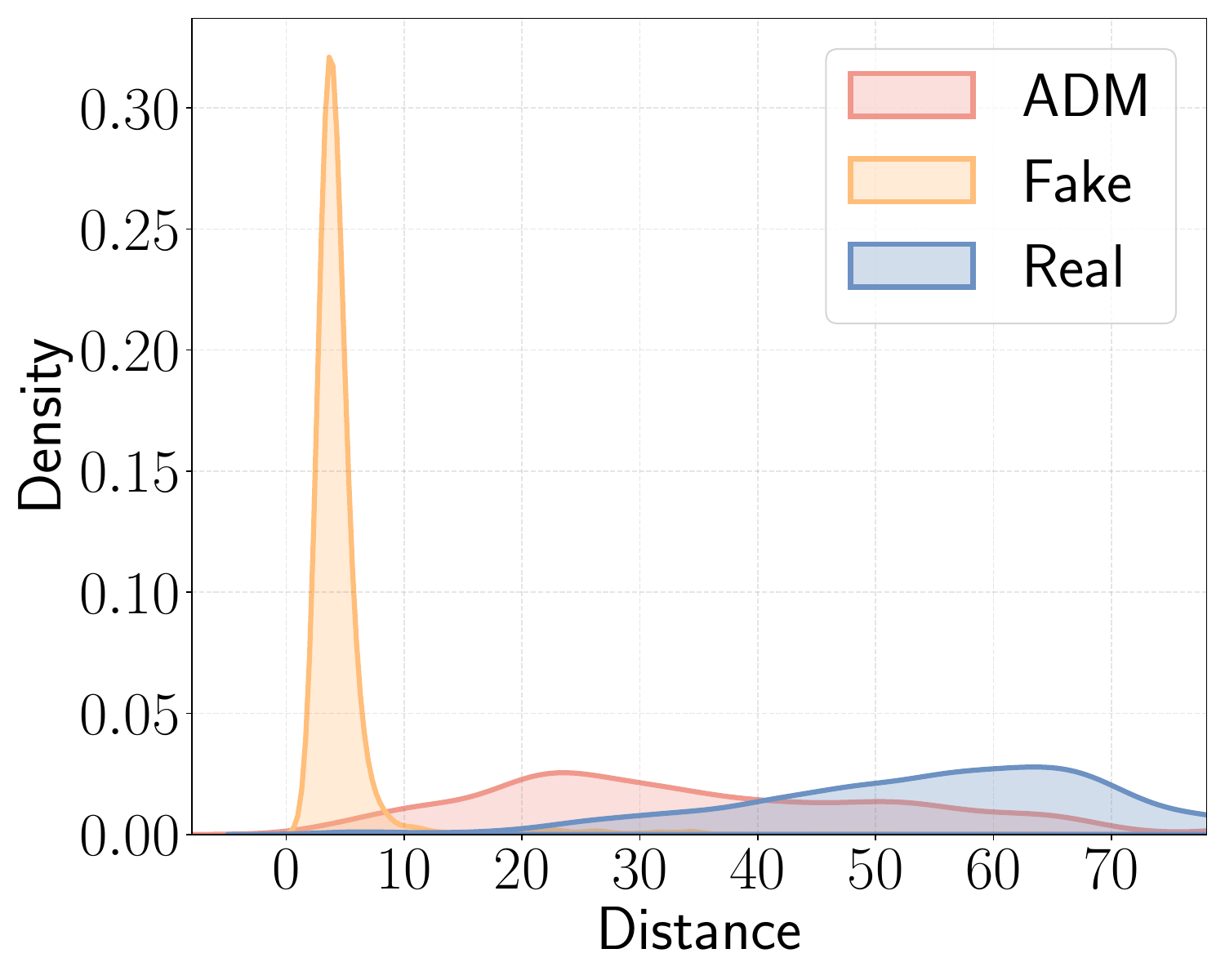}
    \includegraphics[width=0.23\linewidth]{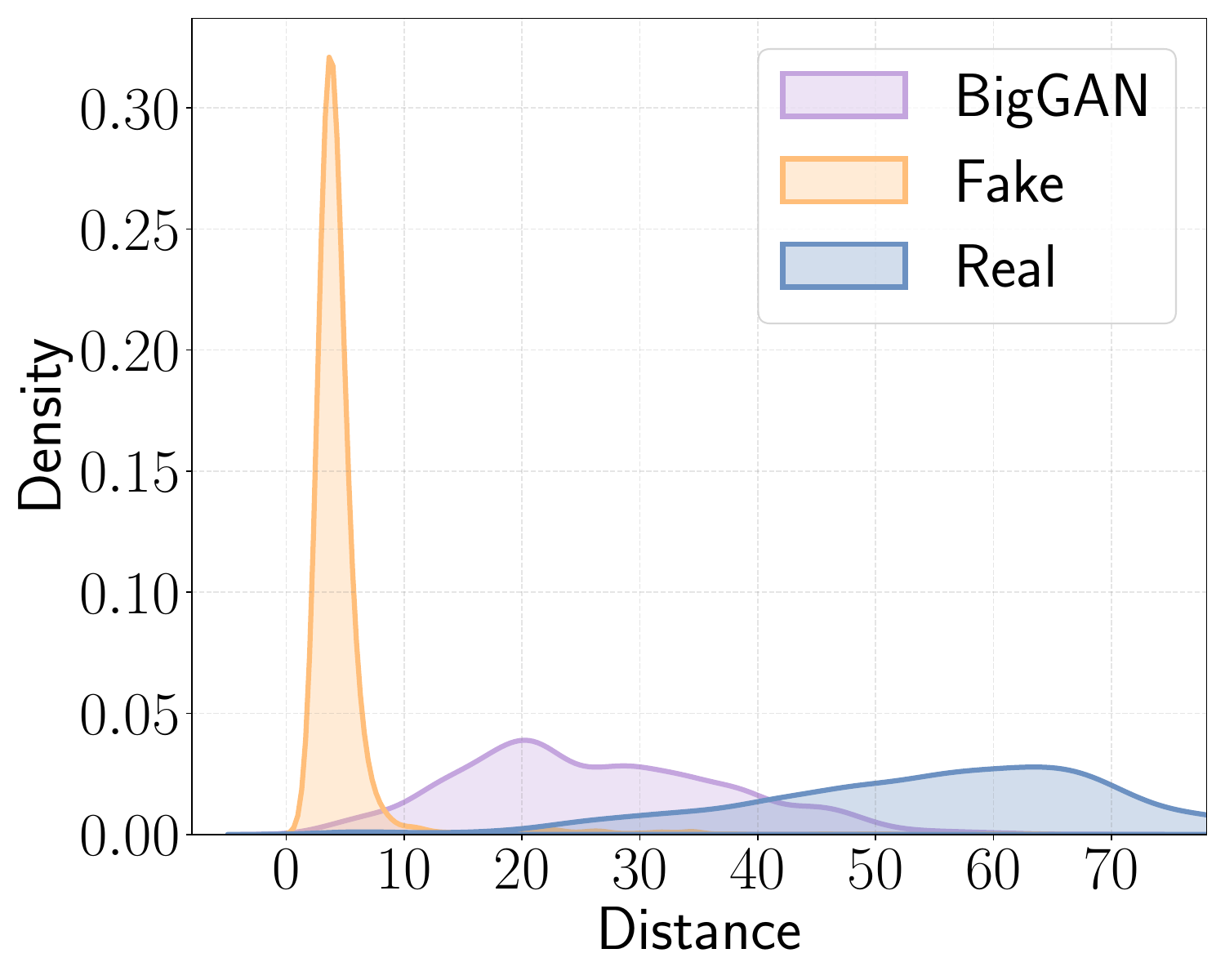} 
    \includegraphics[width=0.23\linewidth]{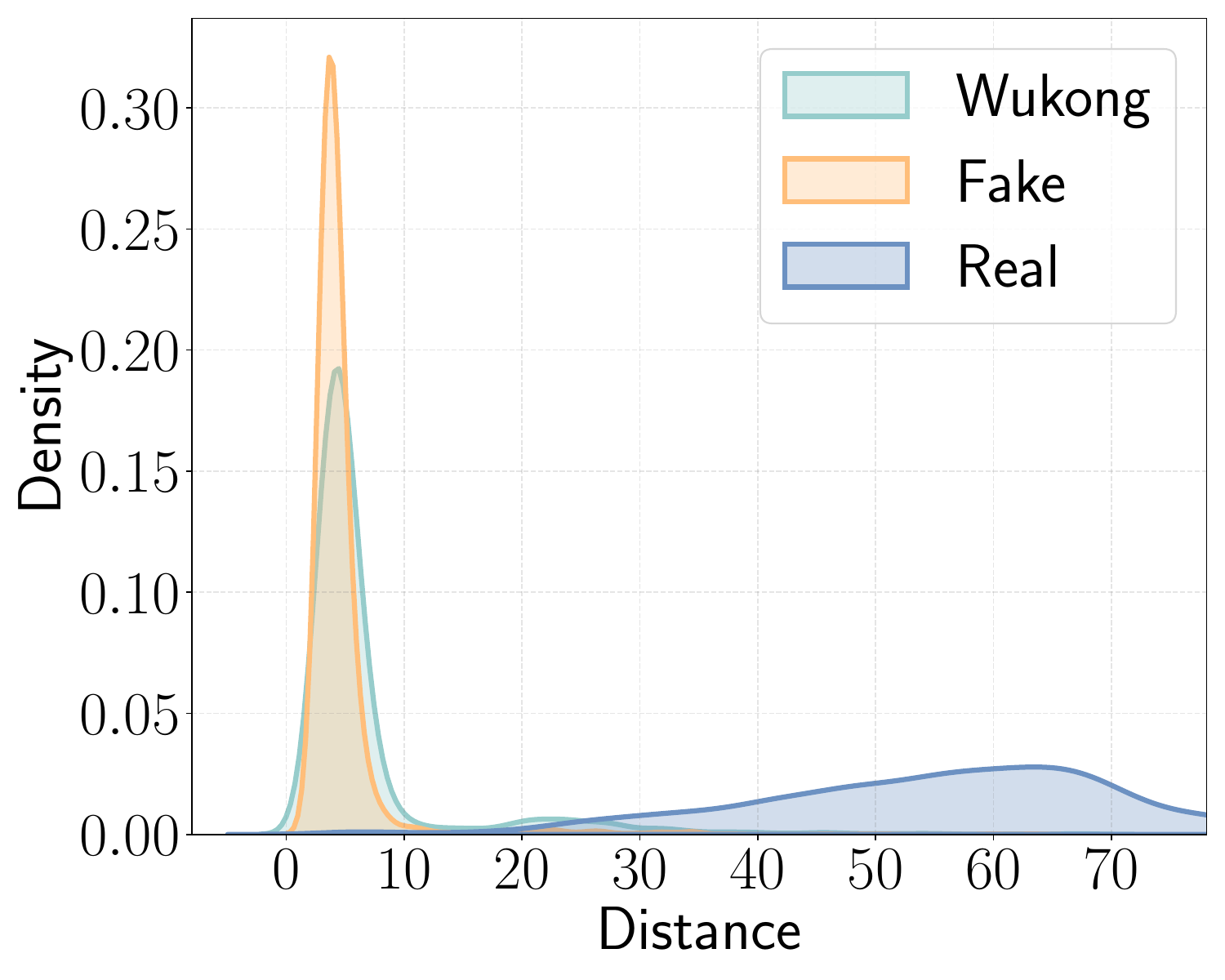} 
  \\
    \includegraphics[width=0.23\linewidth]{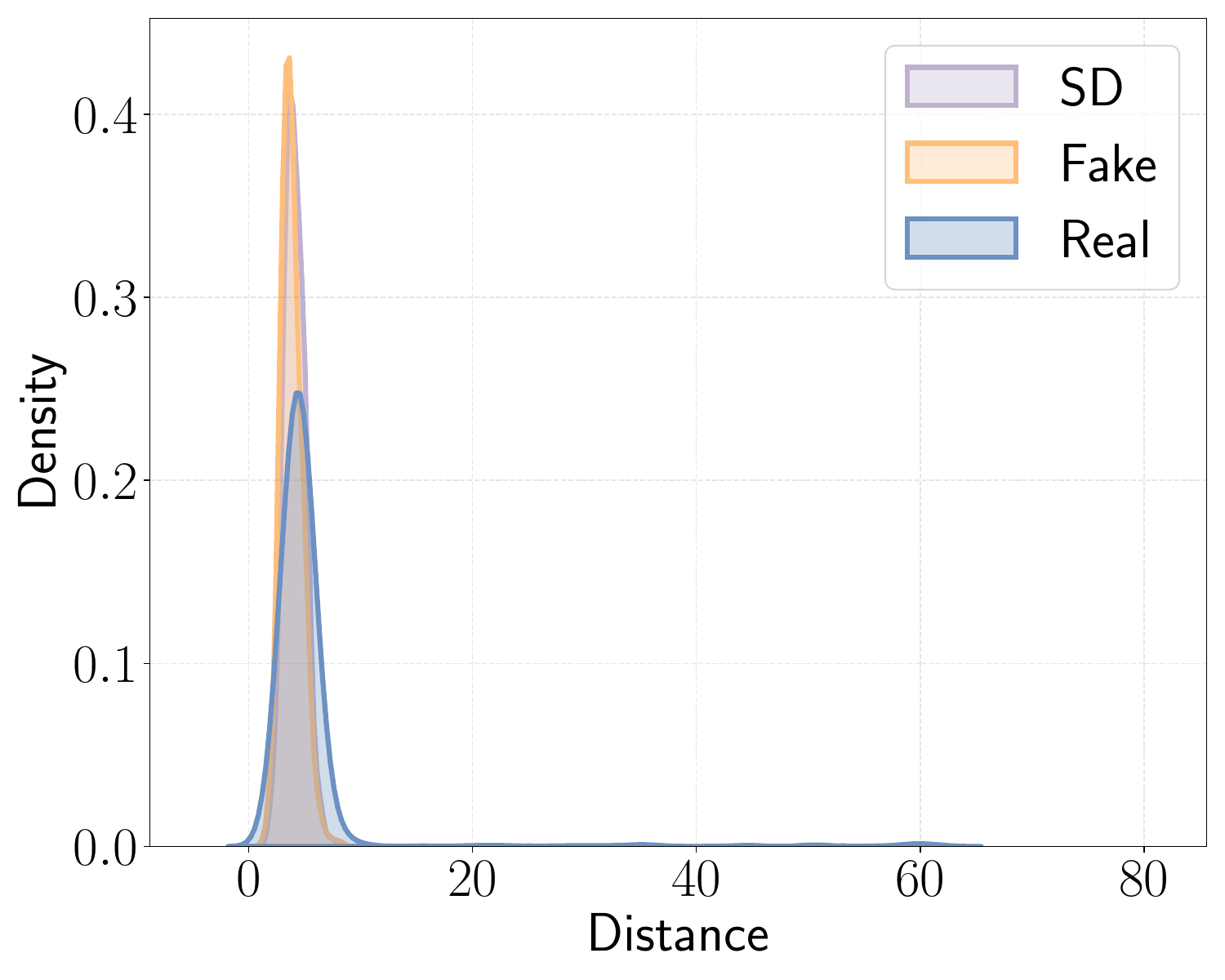} 
    \includegraphics[width=0.23\linewidth]{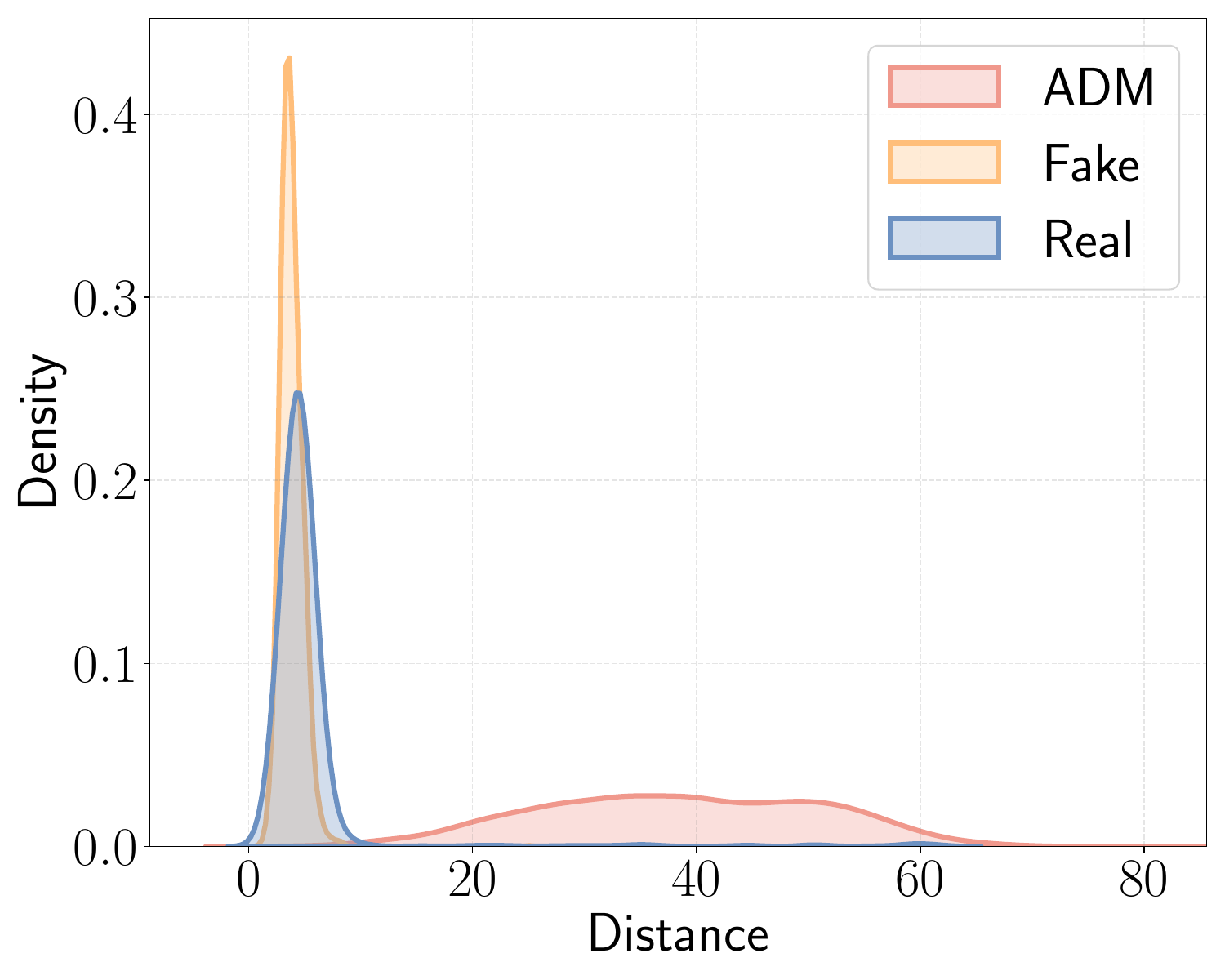}  
    \includegraphics[width=0.23\linewidth]{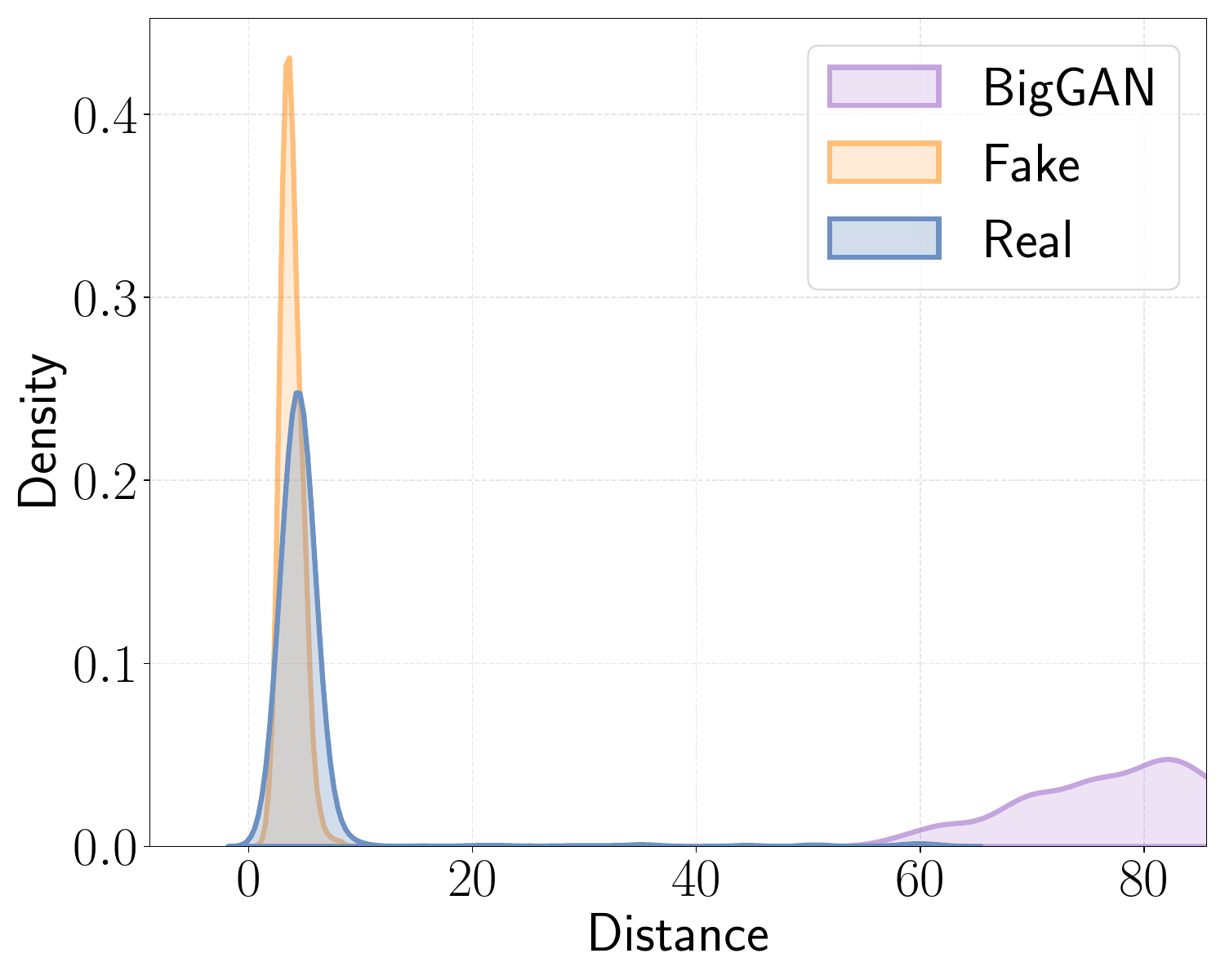} 
    \includegraphics[width=0.23\linewidth]{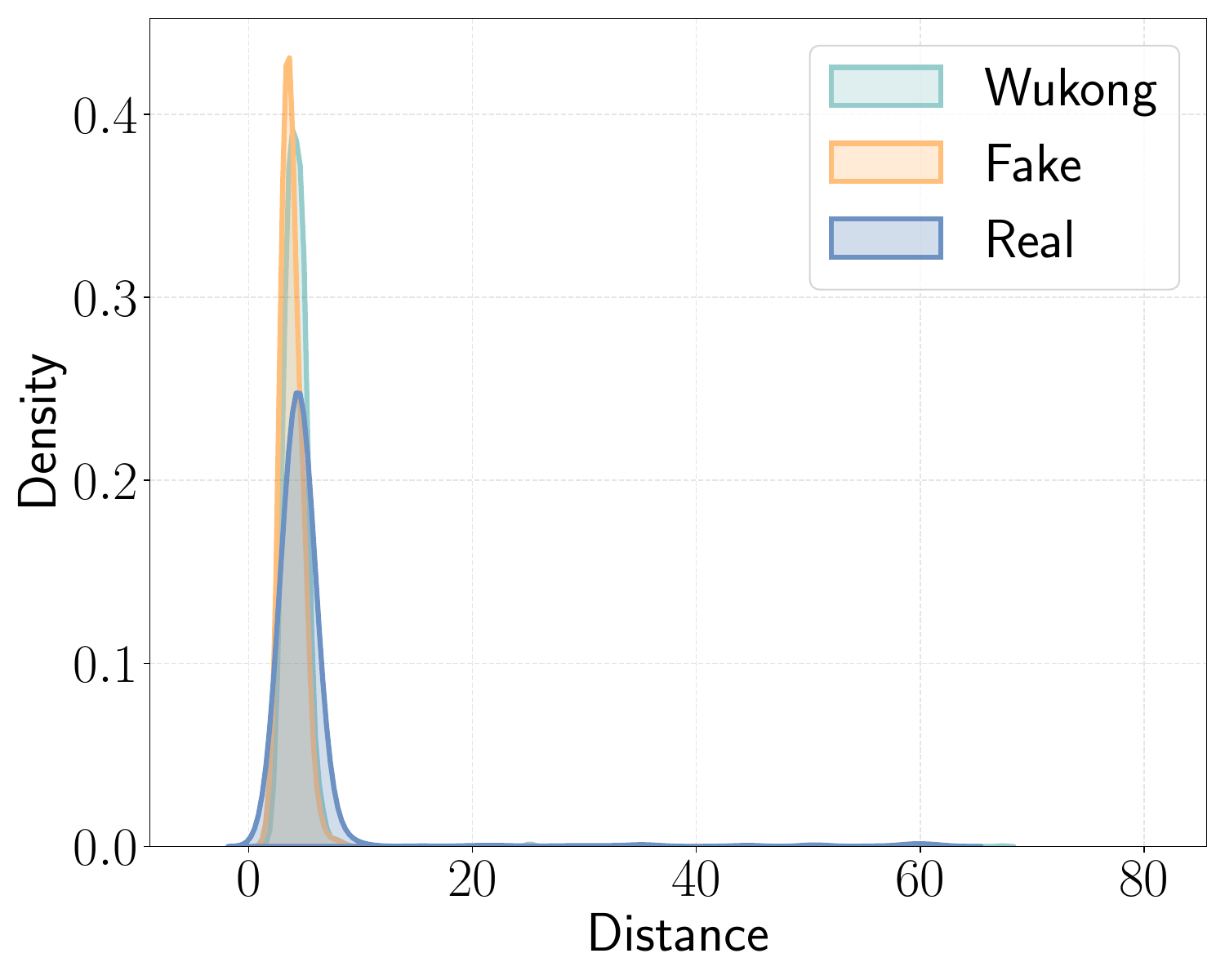} 
  \\
    \caption{KNN distance distributions. Rows correspond to raw and regenerated feature spaces (“Fake” denotes known fake distribution).}
\label{fig:KNN visualization}
\end{figure*}

\subsubsection{Visual Analysis of Distribution Alignment}

In addition to quantitative results, we provide qualitative visualizations to gain insights into how PDA distinguishes real and fake images. Specifically, \autoref{fig:T-SNE visualization} presents t-SNE plots of the feature distributions, and \autoref{fig:KNN visualization} shows the KNN distance distributions. Each column represents a generative model (SD, ADM, BigGAN, and Wukong), and each row corresponds to a specific feature space: the first row represents the raw feature space, and the second row shows the feature space after regeneration using the known generator. Additional results are provided in~\autoref{appendix_visualization}.

\textbf{1) Discriminative decision boundary in the raw feature space.}  
As shown in the first column of the first row (SD), the feature extractor trained on SD effectively captures model-induced artifacts, enabling a clear separation between real and known fake samples in the raw feature space. This confirms that our feature extractor can form a discriminative boundary between real and known fakes, facilitating threshold-based classification.

\textbf{2) Distribution alignment of regenerated real images.}  
The second row demonstrates that real images, when regenerated through the known generator (SD), become pseudo-fakes that inherit the generator's artifact patterns. As seen in the corresponding regenerated space (second row), these pseudo-fakes shift toward the known fake distribution, showing reduced KNN distances. This validates the core mechanism of PDA—mapping real images into the known fake manifold via regeneration to reduce their KNN distances and enable separation from unknown fakes.

\textbf{3) Persistent distributional shift for unknown fakes.}  
The second and third columns (ADM and BigGAN) in the second row show that even after regeneration, unknown fake images continue to exhibit notable distributional shifts and higher KNN distances. These samples fail to align with the known artifact manifold due to their inherited or mixed artifacts from unknown generators. This discrepancy is effectively captured by the KNN-based detector in the second step of PDA.

\textbf{4) Early detection of distribution-similar fakes.}  
Interestingly, the fourth column (Wukong) illustrates a scenario where the unknown fake distribution closely resembles that of the known generator (SD). As shown in the raw feature space (first row), Wukong-generated images are well aligned with the known fake distribution due to similar model-induced artifacts. Consequently, most of these samples are confidently classified as fake in the first step of PDA, without triggering the regeneration process. According to Eq.~\ref{eq:knn3}, only samples predicted as real in the raw space are forwarded to the second step for further analysis. 
This case highlights that PDA dynamically adapts its three-step strategy—applying regeneration based on the separability observed in the raw space. Whether unknown fakes resemble or deviate from the known fake distribution, PDA can accurately classify them.

To further validate our detection mechanism, we analyze the KNN distance distributions in the same t-SNE-reduced feature space. These distances form the basis of our threshold-based decision. As shown in the second and third columns of \autoref{fig:KNN visualization}, real and known fake images exhibit clearly separable distance distributions after regeneration. Pseudo-fake images regenerated from real inputs consistently yield lower distances, aligning closely with the known fake cluster. In contrast, regenerated unknown fakes (e.g., from ADM and BigGAN) retain larger distances due to persistent distributional shifts. These patterns confirm that KNN distance in the learned feature space provides a reliable and interpretable signal for distinguishing real from fake images, and justify its use as the core decision metric in PDA.

These visualizations provide strong empirical support for PDA's theoretical foundation: real images can be aligned with known fakes through regeneration, whereas unknown fakes with distributional shift can be distinguished via KNN-based post-hoc analysis.

\subsection{Ablation Studies}
\label{ablation_studies}
To systematically understand the effectiveness of each component in PDA, we conduct a series of ablation studies. Unless otherwise specified, these experiments are performed using the GenImage dataset~\cite{zhu2024genimage}, which includes real images from ImageNet~\cite{deng2009imagenet} and  synthetic images generated by six representative generative models: Stable Diffusion, GLIDE, VQDM, ADM, BigGAN, and Wukong. For each ablation, we evaluate performance using 3,000 real images and 3,000 synthetic images per model.

\begin{table}[!t]
\centering
\caption{The impact of feature extractor architectures.}
\label{tab:backbone}
\scalebox{0.86}{
\begin{tabular}{lccc}
\toprule
\textbf{Target / Backbone} & \textbf{ResNet-18} & \textbf{ResNet-50} & \textbf{VGG-19} \\
\midrule
SD      & 95.67\% & 96.00\% & 96.93\% \\
GLIDE   & 97.28\% & 98.09\% & 98.43\% \\
VQDM    & 96.89\% & 97.87\% & 98.46\% \\
ADM     & 97.36\% & 98.12\% & 98.46\% \\
BigGAN  & 97.48\% & 98.19\% & 98.46\% \\
Wukong  & 92.35\% & 92.10\% & 94.53\% \\
\midrule
\textbf{AP}    & 96.17\% & 96.73\% & 97.55\% \\
\bottomrule
\end{tabular}}
\end{table}

\subsubsection{Impact of Feature Extractor}
\label{backbone}
To evaluate the generality and compatibility of PDA with different feature extractors, we conduct experiments using three representative network backbones: ResNet-18, ResNet-50~\cite{he2016deep}, and VGG-19~\cite{simonyan2014very}. These architectures vary in depth and capacity, and are widely adopted in existing AI-generated image detectors. \autoref{tab:backbone} reports the detection results across six generative models. Despite architectural differences, PDA consistently achieves strong performance, with AP exceeding 96\% in all cases and reaching up to 97.55\% with VGG-19.

These results demonstrate that PDA is backbone-agnostic and remains effective across a broad range of feature extractors. Unlike conventional detectors that learn fixed decision boundaries from known fakes, PDA avoids explicitly modeling diverse fake distributions. Instead, it aligns real images to the known fake distribution through regeneration, while unknown fakes—containing incompatible or mixed artifacts—remain misaligned, regardless of feature extractor choice. This robustness enables seamless integration with existing detectors and supports practical deployment under open-world generative threats.

\begin{table}[t]
  \centering
  \caption{PDA performance with different known generators.}
  \label{known_generator}
\scalebox{0.86}{
\begin{tabular}{l c c}
\toprule
\textbf{Target /  Known Generator}  & \textbf{SD} & \textbf{Kandinsky 2.2} \\
\midrule
SD      & 96.00\%  & 95.33\%  \\
GLIDE   & 98.09\%  & 96.33\%  \\
VQDM    & 97.87\%  & 95.35\%  \\
ADM     & 98.12\%  & 95.73\%  \\
BigGAN  & 98.19\%  & 95.72\%  \\
Wukong  & 92.10\%  & 95.68\%  \\
\midrule
\textbf{AP}    & 96.73\%  & 95.69\%  \\
\bottomrule
\end{tabular}}
\end{table}

\subsubsection{Impact of Known Generator}
\label{generator}
A key component of our PDA is the known generative model used to regenerate real images and thereby implant consistent, learnable artifacts. While Stable Diffusion (SD)~\cite{rombach2022high} is employed in our primary experiments due to its open-source nature, high quality, and widespread adoption, the effectiveness of PDA is not inherently tied to this specific model. In principle, any generator capable of producing stable and distinctive artifact patterns can serve as the known generator.

To empirically validate the robustness of PDA to the choice of regeneration model, we conducted an ablation study using Kandinsky 2.2~\cite{Kandinsky2.2_ref}, a model inheriting design principles from DALL·E2 ~\cite{ramesh2022hierarchical} and publicly available via HuggingFace. We train feature extractor using Kandinsky-generated images and set the threshold accordingly. During inference, we use Kandinsky for regeneration. As shown in \autoref{known_generator}, PDA achieves a high AP of 95.69\% with Kandinsky 2.2, comparable to 96.73\% with SD, showing the robustness to regeneration model change.

\subsubsection{Impact of Threshold Selection}
\label{threshold}
To assess the robustness of our PDA to its primary hyperparameter, the decision threshold $\tau$, we investigate its impact on detection performance. The threshold $\tau$ is derived from the KNN distance distribution computed between a set of regenerated real images and a distinct reference set of known fake images. Following common practice in out-of-distribution (OOD) detection~\cite{sun2022out,djurisic2022extremely}, $\tau$ is set to a specific percentile of this distribution to control the false positive rate on real samples. For our main experiments, we use the 95th percentile. This threshold is determined using a dedicated set of 3,000 real images, entirely disjoint from any test data, thus ensuring an unbiased evaluation.

We varied the percentile for $\tau$ from the 91th to the 99th percentile and evaluated PDA's performance. As shown in \autoref{threshold_sensitivity}, PDA demonstrates stable performance across a range of reasonable threshold settings. As the percentile varies from 91th to 99th, the AP fluctuates minimally, ranging from 96.53\% to 97.01\%. This insensitivity demonstrates that PDA’s effectiveness stems from robust distribution alignment rather than fine-tuning of threshold parameters, ensuring reliability in open-world detection scenarios.

\begin{table}[t]
  \centering
  \caption{PDA performance with varying ($\tau$). }
  \label{threshold_sensitivity}
  \scalebox{0.86}{
  \begin{tabular}{l c c c c c } 
    \toprule
    \textbf{Target / $\tau$} & \textbf{91th} & \textbf{93th} & \textbf{95th} & \textbf{97th} & \textbf{99th} \\
    \midrule
    SD      & 95.56\%  & 95.83\%  & 96.00\%  & 96.41\%  & 96.63\%  \\
    GLIDE   & 98.04\%  & 98.04\%  & 98.09\%  & 98.24\%  & 97.90\%  \\
    VQDM    & 97.82\%  & 97.82\%  & 97.87\%  & 97.87\%  & 97.49\%  \\
    ADM     & 98.09\%  & 98.09\%  & 98.12\%  & 98.31\%  & 98.12\%  \\
    BigGAN  & 98.15\%  & 98.15\%  & 98.19\%  & 98.46\%  & 98.46\%  \\
    Wukong  & 91.52\%  & 91.52\%  & 92.10\%  & 92.80\%  & 93.21\%  \\
    \midrule
    \textbf{AP} & 96.53\%  & 96.58\%  & 96.73\%  & 97.01\%  & 96.97\%  \\
    \bottomrule
  \end{tabular}}
\end{table}

\begin{figure}[!t]
    \centering
    \includegraphics[width=0.36\textwidth]{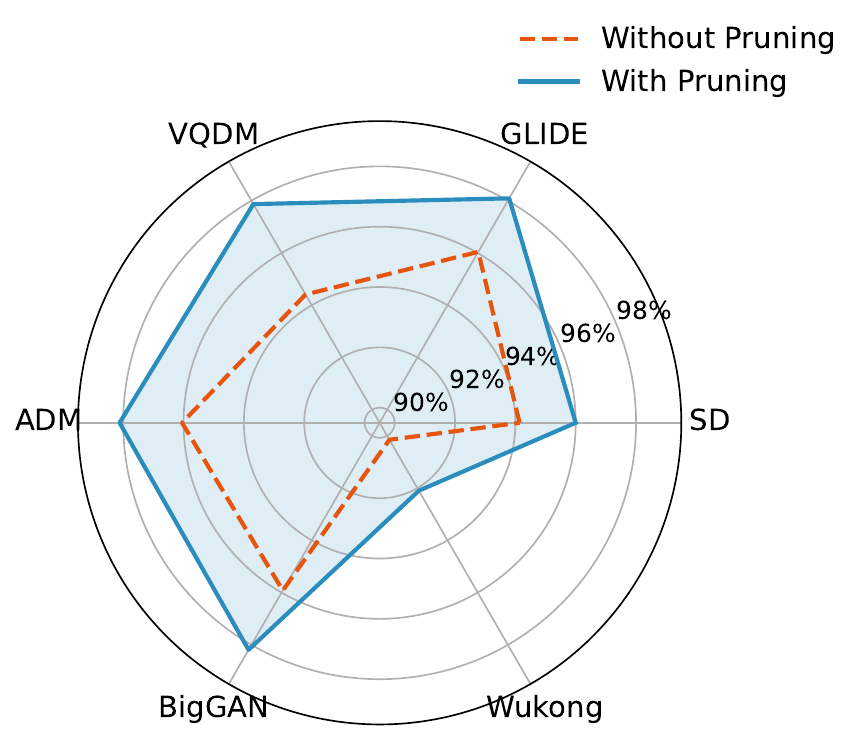}
    \caption{The impact of activation pruning.}
    \label{fig:activate}
\end{figure}

\subsubsection{Impact of Activation Pruning}
We evaluate the effect of activation pruning on PDA’s detection performance. Prior studies~\cite{sun2021react,djurisic2022extremely} suggest that out-of-distribution (OOD) inputs often trigger abnormally high activations in specific feature dimensions, which can degrade the reliability of downstream classification.

To address this, we adopt activation pruning to suppress excessive activations before applying our detection mechanism. As shown in \autoref{fig:activate}, pruning consistently improves detection performance across all tested generative models. This enhancement is attributed to the mitigation of spurious high responses, which helps refine the feature space and leads to more robust and stable decision boundaries under distribution shift. These results highlight the effectiveness of activation rectification as a lightweight and generalizable enhancement to improve the resilience of PDA against diverse unknown fakes.

\begin{table}[!t]
  \centering
  \caption{PDA performance with different reduction tools.}
  \label{tab:tsne_pca}
  \scalebox{0.88}{
  \begin{tabular}{l c c}
  \toprule
  \textbf{Target / Reduction Tool} & \textbf{PCA} & \textbf{t-SNE (Ours)} \\
  \midrule
  SD & 94.85\% & 96.00\% \\
  GLIDE            & 95.35\% & 98.09\% \\
  VQDM             & 95.63\% & 97.87\% \\
  ADM              & 96.18\% & 98.12\% \\
  BigGAN           & 96.82\% & 98.19\% \\
  Wukong           & 92.10\% & 92.10\% \\
  \midrule
  \textbf{AP}      & 95.66\% & 96.73\% \\
  \bottomrule
  \end{tabular}}
\end{table}

\begin{figure}[!t]
    \centering
    \includegraphics[width=0.42\textwidth]{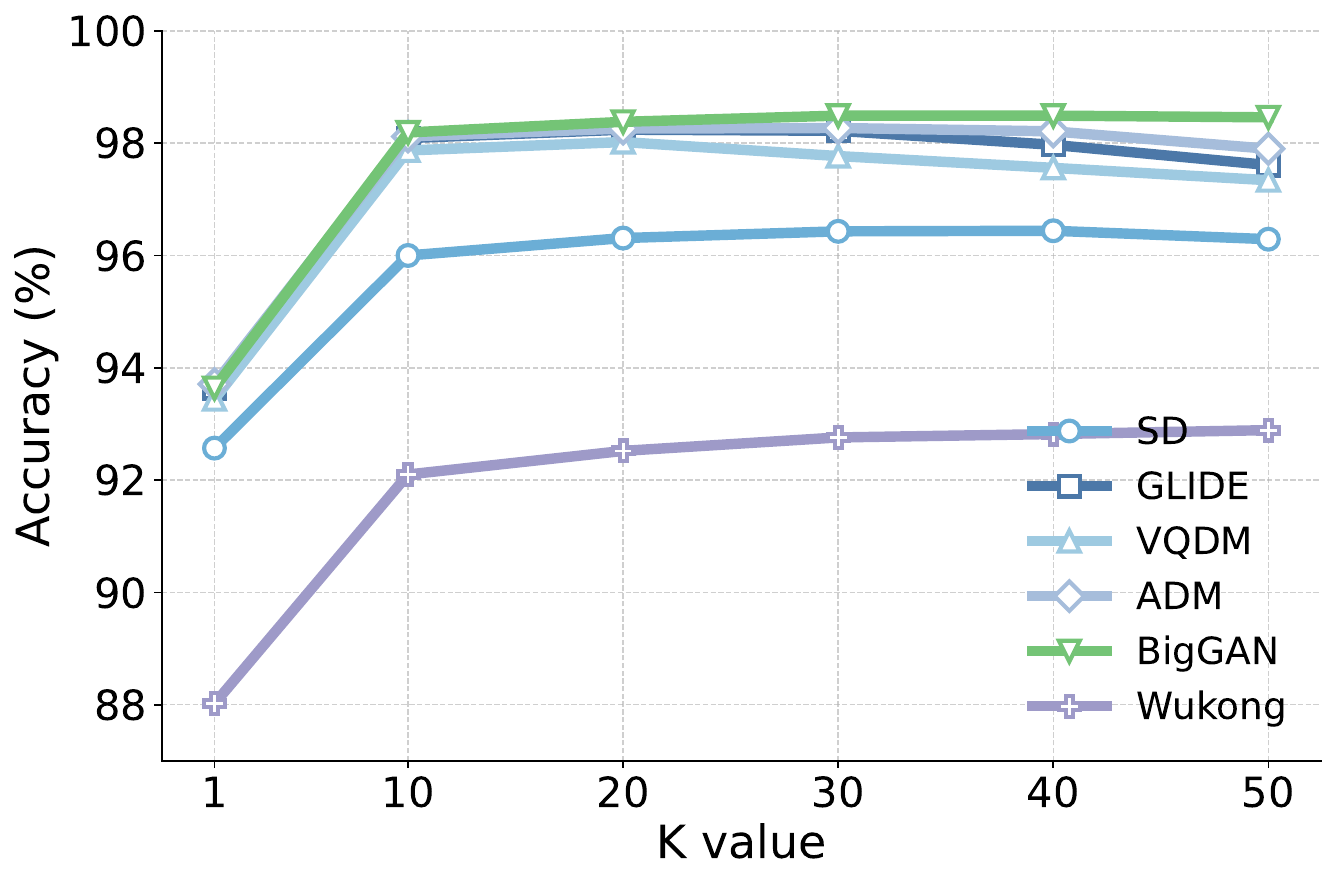}
    \caption{Detection accuracy of PDA across different K values.}
    \label{fig:knn_k_curve}
\end{figure}

\subsubsection{Impact of Dimensionality Reduction}
\label{ablation_tsne}
PDA employs t-SNE~\cite{van2008visualizing} to project pruned feature vectors into 2D for efficient KNN computation while preserving local geometry. To evaluate robustness, we replace t-SNE with PCA and re-evaluate detection across all generators. As shown in \autoref{tab:tsne_pca}, PDA achieves comparable performance with both methods (e.g., 98.19\% vs. 96.82\% on BigGAN), with negligible differences for other generators such as Wukong. These results confirm that PDA’s effectiveness derives from its distribution alignment mechanism rather than dependence on a specific dimensionality reduction strategy.

\subsubsection{Impact of KNN}
\label{ablation_knn}
We systematically analyze the effect of the number of neighbors \( k \) on the performance of our KNN-based detection approach. In this experiment, we vary \( k \) across the values \( \{1, 10, 20, 30, 40, 50\} \) and evaluate the performance of the method in terms of detection accuracy. 

\begin{figure}[!t]
    \centering
    \includegraphics[width=0.38\textwidth]{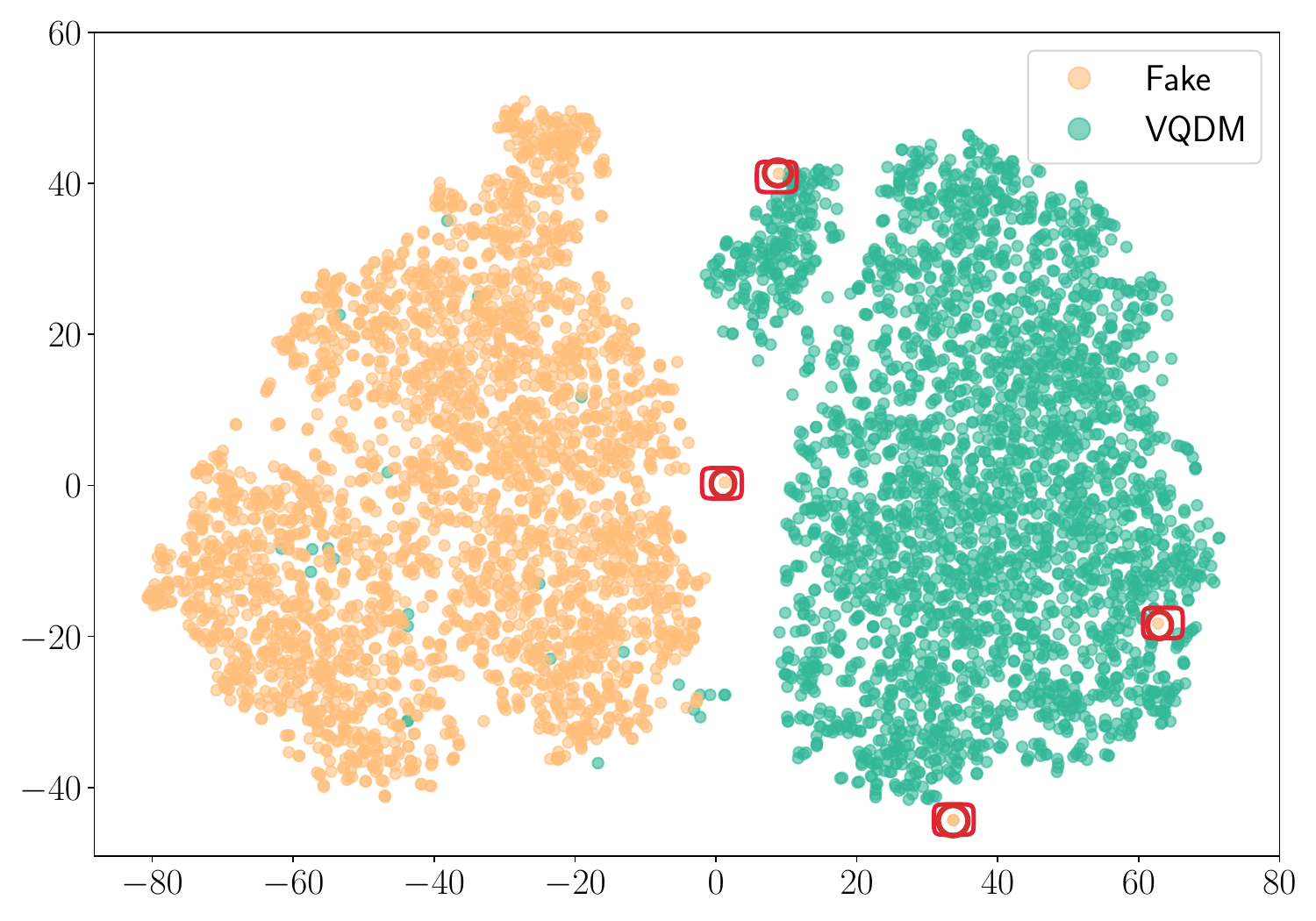}
    \caption{Outliers analysis in feature space.}
    \label{fig:outlies}
\end{figure}

As shown in \autoref{fig:knn_k_curve}, our method achieves superior performance and remains stable across different values of \( k \), except when \( k = 1 \). The performance for other values of \( k \) shows consistent results and has little impact on the detection performance across different generative models. In \autoref{fig:outlies}, we demonstrate why the performance is poor when \( k = 1 \). Specifically, the nearest neighbor reference set may contain outliers, such as the red-circled data points that fall outside the distribution (anomalous points). These outliers can affect the KNN distance distribution, causing fake samples (e.g., those generated by VQDM) to have small distances to the nearest neighbors after regeneration, making it difficult to distinguish them from real samples. By using a larger \( k = 10 \) (or other values), we effectively mitigate the impact of these outliers, ensuring robust and reliable detection.

\subsubsection{Impact of Training Dataset Size}
\label{ablation_size}
We analyze the effect of training dataset size used to train the detector feature extractor on the performance of PDA. 
Specifically, we vary the number of fake training samples from 20,000 to 120,000, using an equal number of real images in each setting to maintain class balance.

\begin{figure}[!t]
    \centering
    \includegraphics[width=0.45\textwidth]{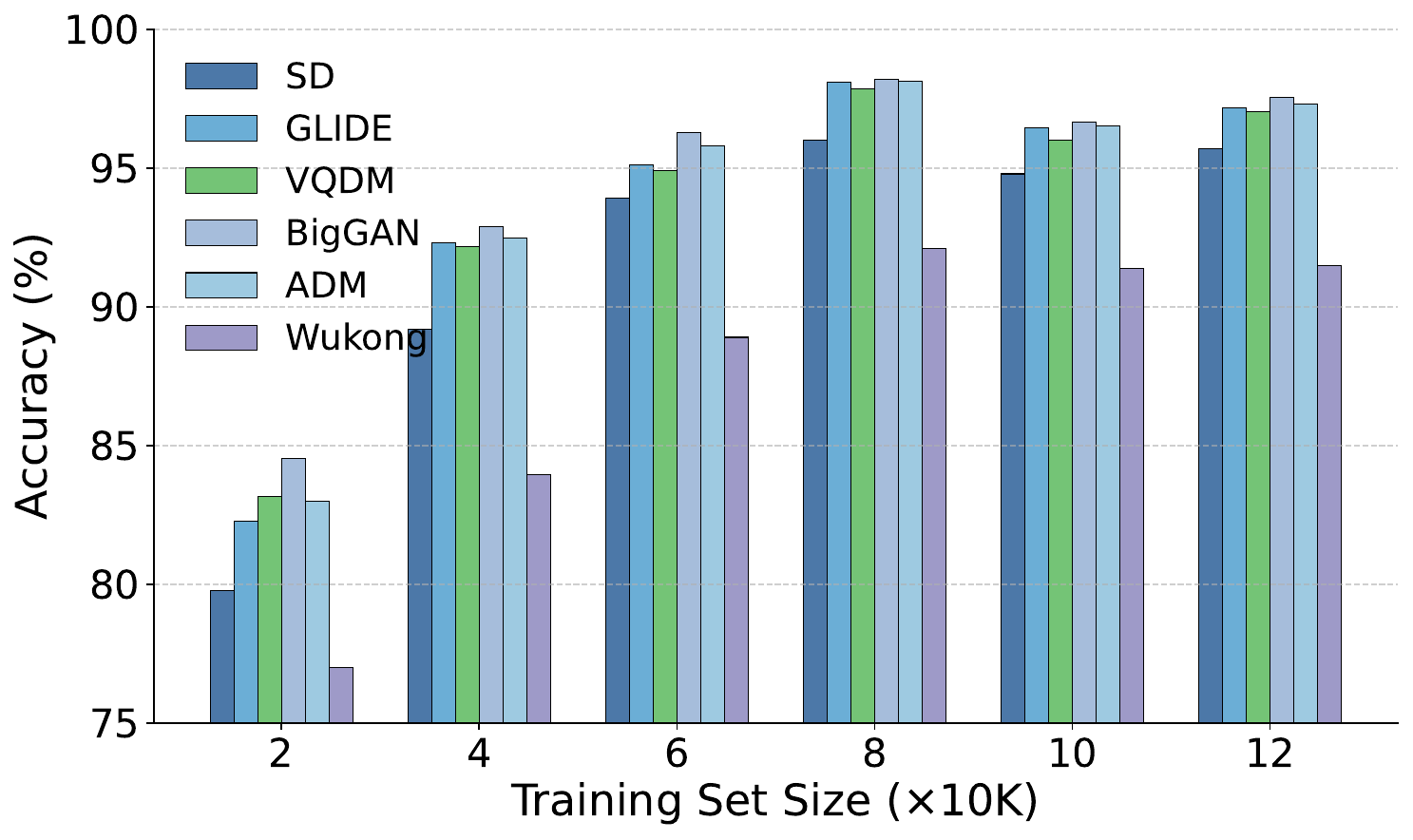}
    \caption{Detection accuracy on different training set sizes.}
    \label{fig:training_size_bar}
\end{figure}

\autoref{fig:training_size_bar} indicates that while performance improves significantly in the low-data regime (81.6\% to 94.15\% AP), gains marginalize beyond 60k samples. 
Notably, this analysis concerns only the training of feature extractor, while PDA’s inference-time alignment and detection procedure remains training-free. 
These results indicate that PDA achieves strong performance with moderate training data, underscoring its data efficiency and practical scalability in settings with limited annotations. 


\begin{figure*}[!t]
\centering
\begin{minipage}{0.13\linewidth}
    \centerline{\includegraphics[width=\textwidth]{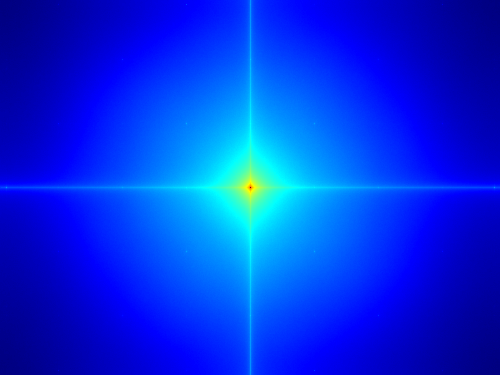}}
    \vspace{6pt}
    \centerline{\includegraphics[width=\textwidth]{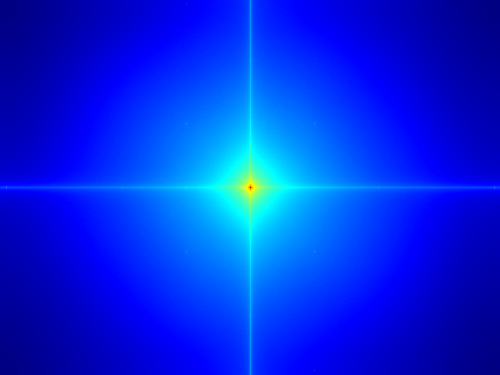}}
    \centerline{SD}
\end{minipage}
\begin{minipage}{0.13\linewidth}
    \centerline{\includegraphics[width=\textwidth]{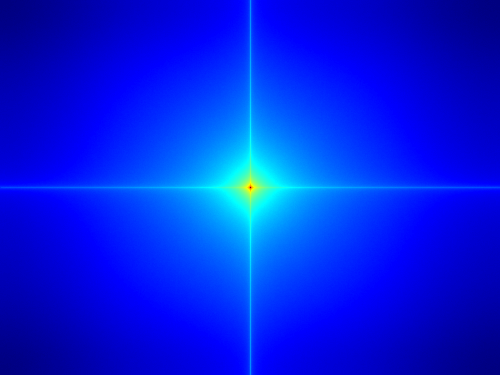}}
    \vspace{6pt}
    \centerline{\includegraphics[width=\textwidth]{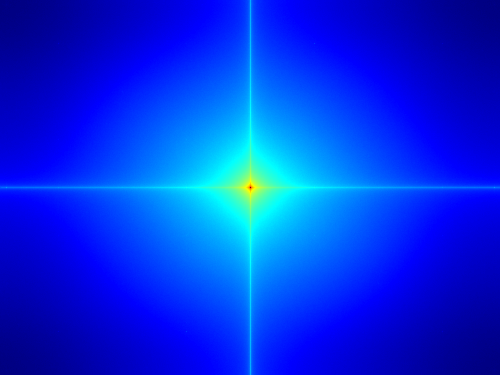}}
    \centerline{Real}
\end{minipage}
\begin{minipage}{0.13\linewidth}
    \centerline{\includegraphics[width=\textwidth]{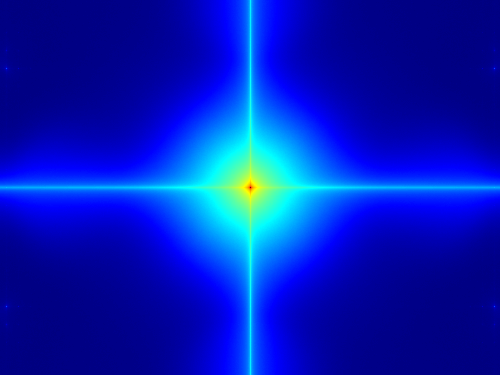}}
    \vspace{6pt}
    \centerline{\includegraphics[width=\textwidth]{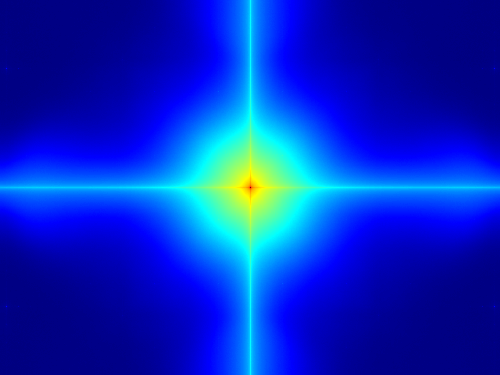}}
    \centerline{GLIDE}
\end{minipage}
\begin{minipage}{0.13\linewidth}
    \centerline{\includegraphics[width=\textwidth]{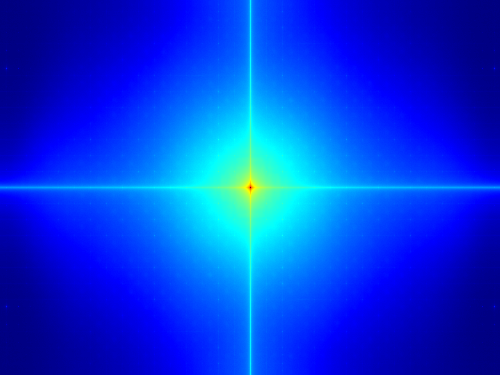}}
    \vspace{6pt}
    \centerline{\includegraphics[width=\textwidth]{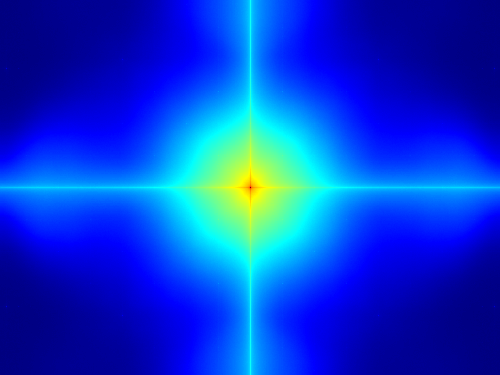}}
    \centerline{VQDM}
\end{minipage}
\begin{minipage}{0.13\linewidth}
    \centerline{\includegraphics[width=\textwidth]{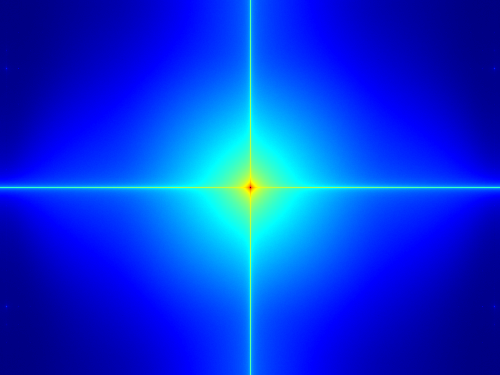}}
    \vspace{6pt}
    \centerline{\includegraphics[width=\textwidth]{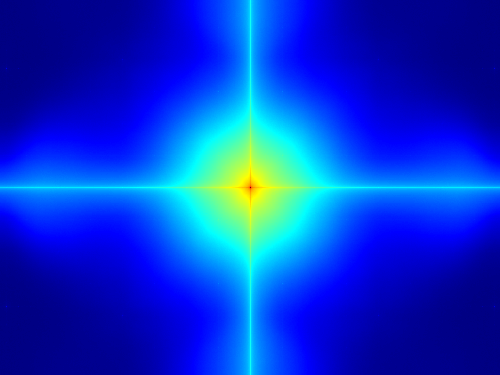}}
    \centerline{ADM}
\end{minipage}
\begin{minipage}{0.13\linewidth}
    \centerline{\includegraphics[width=\textwidth]{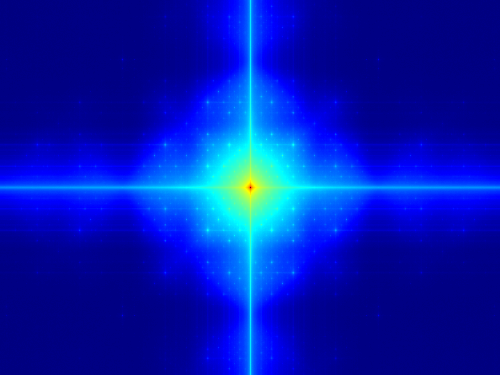}}
    \vspace{6pt}
    \centerline{\includegraphics[width=\textwidth]{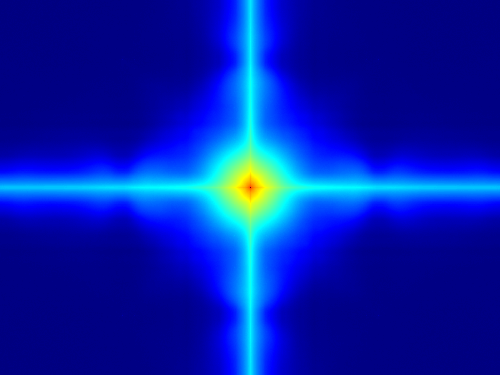}}
    \centerline{BigGAN}
\end{minipage}
\begin{minipage}{0.13\linewidth}
    \centerline{\includegraphics[width=\textwidth]{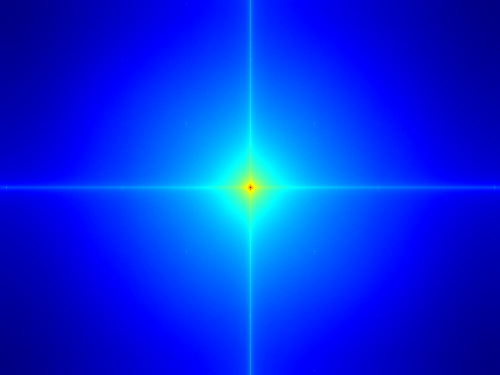}}
    \vspace{6pt}
    \centerline{\includegraphics[width=\textwidth]{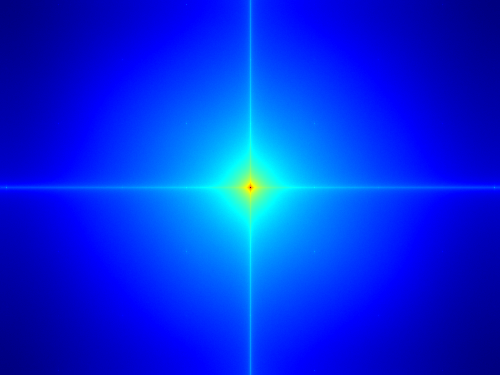}}
    \centerline{Wukong}
\end{minipage}
\caption{Visualization of frequency spectra for original (first row) and regenerated images using the known generator (SD).}
\label{fig:frequency}
\end{figure*}

\subsection{Frequency Analysis}
To further validate PDA’s design, we analyze frequency-domain patterns of real and fake images by computing the average Fourier spectra~\cite{sha2023fake}, as shown in \autoref{fig:frequency}. Real images exhibit smooth low-frequency distributions, while fakes from models such as BigGAN, ADM, and GLIDE show irregular high-frequency components, revealing generator-specific artifacts. Notably, Wukong shares similar spectra with SD, explaining its easier detection in PDA’s first stage. 

After regeneration with SD, real images acquire spectral patterns nearly identical to SD-generated fakes, confirming alignment with the known fake distribution. In contrast, regenerated unknown fakes (e.g., ADM, GLIDE, BigGAN) retain noticeable frequency discrepancies, reflecting inconsistent artifacts between unknown and known generators. These findings provide complementary evidence for PDA: regeneration consistently implants stable, learnable artifacts into real images, while unknown fakes remain misaligned in both spectral and feature spaces, thereby enabling robust detection.

\begin{table}[!t]
\centering
\caption{Generalization across diverse real-image datasets.}
\label{tab:cross_distribution}
\scalebox{0.85}{
\begin{tabular}{l|cccc|c}
\toprule
\textbf{Dataset} & ImageNet & LSUN & COCO & CelebA-HQ & AP \\
\midrule
\textbf{PDA} & 96.87\% & 95.78\% & 96.70\% & 98.60\% & 96.99\% \\
\bottomrule
\end{tabular}}
\end{table}

\subsection{Generalization Across Real Images}
\label{Performance_real}
To assess the robustness of PDA under varying real-world content distributions, we evaluate its performance on real images drawn from four representative datasets: ImageNet~\cite{deng2009imagenet}, LSUN~\cite{yu2015lsun}, COCO~\cite{lin2014microsoft}, and CelebA-HQ~\cite{karras2017progressive}. These datasets cover a wide semantic and distributional range, posing a strong test for cross-distribution generalization.

As shown in \autoref{tab:cross_distribution}, PDA maintains consistently high accuracy across all datasets, achieving an AP of 96.99\%. This strong generalization is attributed to the design of PDA: by regenerating real images through a fixed known generative model, consistent model-specific artifacts are introduced regardless of the original image semantics. These artifacts, already captured by the feature extractor trained on known fakes, lead to effective post-hoc alignment between regenerated real images and known fake distributions. Thus, PDA does not rely on the semantic content of real images and remains robust even when applied to data with unknown or diverse real-world distributions.

\subsection{Inference Efficiency}
\label{runtime}
We evaluate the inference efficiency of PDA by measuring the average runtime per 100 samples on an NVIDIA RTX 3090 (24GB VRAM). The reported time covers both regeneration with a known generator and subsequent KNN-based classification. For comparison, we focus on ZeroFake~\cite{sha2024zerofake}, the most relevant training-free baseline, since it also performs a reconstruction step during inference and has been shown to outperform earlier methods such as DIRE~\cite{wang2023dire}.

As shown in \autoref{tab:inference_time}, PDA requires only 4.06 seconds, which is nearly 37× faster than ZeroFake (149.18 seconds). The efficiency gain comes from PDA’s lightweight pipeline: KNN classification step is negligible, requiring only a forward pass and distance computation against a fixed 3,000-sample reference set. The main computational cost comes from the regeneration step, which is still significantly cheaper than diffusion-based reconstruction and adversarial optimization used in ZeroFake. These results highlight that PDA not only achieves strong detection accuracy but also offers practical efficiency for latency-sensitive and resource-constrained deployment scenarios.

\begin{table}[!t]
\centering
\caption{Average inference time for 100 samples.}
\label{tab:inference_time}
\scalebox{1}{ 
\begin{tabular}{l|c}
\toprule
\textbf{Method} & \textbf{Time (s)} \\
\midrule
ZeroFake~\cite{sha2024zerofake} & 149.18 \\
\textbf{PDA (Ours)} & \textbf{4.06} \\
\quad -- Regeneration & 3.58 \\
\quad -- KNN Classification & 0.48 \\
\bottomrule
\end{tabular}}
\end{table}

\begin{table*}[!t]
  \centering
  \caption{Robustness of PDA detection under diverse image transformations. }
  \scalebox{0.92}{  
    \begin{tabular}{c|c|ccccccc}
    \toprule
    \textbf{Transformation} & \textbf{Factor} & \textbf{SD} & \textbf{GLIDE} & \textbf{VQDM} & \textbf{ADM} & \textbf{BigGAN} & \textbf{Wukong} & \textbf{AP} \\
    \midrule
    \multirow{3}[2]{*}{Gaussian Blurring } & Kernel Size = 3 & 95.73\% & 97.38\% & 97.10\% & 97.57\% & 97.69\% & 92.13\% & 96.27\% \\
          & Kernel Size = 5 & 95.90\% & 97.29\% & 97.24\% & 97.61\% & 97.77\% & 93.88\% & 96.62\% \\
          & Kernel Size = 7 & 95.76\% & 97.17\% & 96.87\% & 97.29\% & 97.56\% & 92.76\% & 96.24\% \\
    \midrule
    \multirow{3}[2]{*}{Image Compression} & QF = 90 & 96.17\% & 98.13\% & 97.60\% & 98.20\% & 98.27\% & 92.25\% & 96.77\% \\
          & QF = 70 & 96.33\% & 98.25\% & 97.91\% & 98.33\% & 98.40\% & 93.80\% & 97.17\% \\
          & QF = 50 & 96.06\% & 98.06\% & 97.71\% & 98.16\% & 98.25\% & 92.20\% & 96.74\% \\
    \midrule
    \textbf{PDA} & No Transformation   & 96.00\% & 98.09\% & 97.87\% & 98.12\% &98.19\% & 92.10\% & 96.73\% \\
    \bottomrule
    \end{tabular}%
  \label{tab:robustness}
  }
\end{table*}%

\subsection{Robustness Evaluation}

We further evaluate the robustness of PDA under common image transformations with varying parameters, including:
(1) \textbf{Gaussian blurring}~\cite{kurakin2018adversarial}, applied with different kernel sizes; and
(2) \textbf{image compression}, evaluated using different quality factors (QF), where lower QF indicates stronger compression.

As shown in \autoref{tab:robustness}, our PDA demonstrates strong robustness across various image transformations. The performance, in terms of ACC and AP, remains high even under challenging conditions. Notably, PDA achieves AP scores of over 96\% despite the introduction of noise and image distortions. These results highlight the effectiveness and resilience of the PDA method in real-world applications, where images are often subject to various image transformations.


\section{Conclusion \& Future Work}

We propose \textit{Post-hoc Distribution Alignment (PDA)}, a generalized and model-agnostic framework for detecting AI-generated images under open-world generative threats. PDA reformulates detection as a distribution alignment task: real images are regenerated through a known generator to align with its distribution, while fake images generated by previously unknown models remain misaligned. This simple yet effective design enables detectors trained on one known generator to generalize to unknown fakes without retraining. Extensive experiments across 16 diverse generative models—including GANs, diffusion models, and commercial text-to-image APIs—demonstrate that PDA achieves state-of-the-art performance (96.69\% average accuracy), while maintaining robustness under distribution shifts and image transformations. These results highlight PDA’s scalability and practicality for open-world AI-generated image detection. 

PDA provides three key advantages over prior approaches: (i) a \textit{model-agnostic formulation}  that integrates seamlessly with existing detection architectures, (ii) \textit{training-free generalization} that enables reliable detection of unknown fakes without retraining, and (iii) \textit{computational efficiency} achieved through a single regeneration step, avoiding iterative optimization or test-time adaptation. In future work, we will examine adversarial threats to PDA’s alignment mechanism, explore its integration with authenticity infrastructures (e.g., watermarking, provenance tracking), and extend PDA to other generative modalities such as AI-generated videos, broadening its applicability in open-world generative ecosystems. 


\section*{Ethical Considerations}

This work studies the detection of AI-generated images under open-world generative threats, with the primary goal of improving the reliability of content authenticity verification systems. Our proposed method, \textit{Post-hoc Distribution Alignment (PDA)}, is designed as a defensive technique to help mitigate the misuse of generative models in applications such as misinformation dissemination, identity fraud, and content manipulation.

\mypara{Potential Risks}
As with many detection techniques, there is a risk that insights from this work could be misused to inform adversaries about potential weaknesses of existing detectors or to guide the design of more evasive generative models. Additionally, incorrect deployment or overreliance on automated detection systems may lead to false positives or negatives, which could have downstream consequences in high-stakes applications such as content moderation or identity verification.

\mypara{Risk Mitigation}
We take several steps to minimize potential harm. First, our work focuses on detection rather than generation, and does not introduce new techniques for creating or enhancing deceptive content. Second, the proposed framework operates at a high level of abstraction and does not rely on exploiting specific vulnerabilities of individual detectors or platforms. Third, all experiments are conducted on publicly available datasets and models, without involving human subjects, private user data, or sensitive personal information.

\mypara{Responsible Use}
We emphasize that PDA is intended to complement, rather than replace, human judgment and existing safeguards in real-world systems. Practitioners deploying this method should carefully consider application-specific thresholds, error tolerances, and the broader socio-technical context in which automated detection is used. We encourage future work to further examine the societal impacts of AI-generated content detection and to develop best practices for responsible deployment.

\cleardoublepage

\bibliographystyle{plain}
\bibliography{sample-base1}

\cleardoublepage

\appendix


\section{Description of Generation Models}
\label{appendix_models}
We provide detailed descriptions of the generative models evaluated in our experiments. These models represent a variety of approaches to image generation, including GAN, diffusion-based models, and text-to-image models.

\begin{itemize}
    \item \textbf{ProGAN~\cite{karras2017progressive}.} A generative adversarial network that introduces a progressive training methodology. It starts from a low-resolution image and incrementally adds new layers to model finer details, significantly improving both generation quality and training stability for high-resolution image synthesis.

    \item \textbf{StyleGAN~\cite{karras2019style}.} An advanced GAN architecture that introduces a style-based generator. It allows for intuitive, scale-specific control over image synthesis by disentangling high-level attributes (e.g., pose, identity) from stochastic variation (e.g., hair, freckles) and enables convincing non-linear interpolation.

    \item \textbf{BigGAN~\cite{brock2018large}.}  A large-scale GAN model that generates high-resolution images by leveraging a combination of class-conditional batch normalization and orthogonal regularization. BigGAN is known for its ability to produce diverse and realistic images across multiple categories. 
    
    \item \textbf{CycleGAN~\cite{zhu2017unpaired}.} An image-to-image translation model that learns to map images from a source domain to a target domain without paired training examples. It utilizes a cycle-consistency loss to ensure that if an image is translated to the target domain and back, it should resemble the original image.
    
    \item \textbf{StarGAN~\cite{choi2018stargan}.} A unified GAN framework capable of performing multi-domain image-to-image translation using a single generator and discriminator. It learns mappings between multiple domains by conditioning the generator with a target domain label, enabling flexible translation across diverse attributes.
    
    \item \textbf{GauGAN~\cite{park2019semantic}.} A semantic image synthesis model that generates photorealistic images from semantic layout masks. It employs Spatially-Adaptive Normalization (SPDAE) layers that modulate activations using semantic masks, allowing for fine-grained control over the style and content.
    
    \item \textbf{StyleGAN2~\cite{karras2020analyzing}.} An improved version of StyleGAN that addresses several characteristic artifacts (e.g., water-droplet artifacts) by redesigning generator normalization, progressive growing, and regularization. It achieves state-of-the-art results in unconditional image synthesis with enhanced image quality and better disentanglement.
    
    \item \textbf{WhichFaceIsReal~\cite{whichfaceisreal_dataset}.} A benchmark dataset and online platform designed for evaluating the detection of AI-generated faces, typically produced by advanced GANs like StyleGAN~\cite{karras2019style}. It highlights the challenge of distinguishing highly realistic synthetic faces from real ones.
    
    \item \textbf{Midjourney~\cite{midjourney_website}.} A commercial text-to-image generation service known for producing artistic and often surreal high-quality images from textual prompts. It operates as a closed-source model accessible via an API, popular for its distinctive aesthetic style and ease of use.
    
    \item \textbf{DALL·E 2~\cite{ramesh2022hierarchical}.} A powerful text-conditional image generation system from OpenAI that can create realistic images and art from natural language descriptions. It utilizes a diffusion model conditioned on CLIP image latents, demonstrating capabilities in generating diverse outputs, inpainting, and variations of existing images.

    \item \textbf{SDXL~\cite{podell2023sdxl}.} An advanced latent diffusion model designed for high-resolution text-to-image synthesis, representing a significant improvement over earlier Stable Diffusion models. It features a larger U-Net backbone and a refined conditioning scheme, enabling the generation of more detailed and aesthetically pleasing images, particularly at 1024$\times$1024 resolution.

    \item \textbf{Stable Diffusion V1.4~\cite{rombach2022high}.} A latent diffusion model that generates high-quality images by iteratively denoising a random latent vector. It is trained on large-scale datasets and is known for its ability to produce highly realistic images with fine details.
    
    \item \textbf{GLIDE~\cite{nichol2021glide}.} A text-to-image diffusion model that leverages guided diffusion to generate images conditioned on textual descriptions. GLIDE is notable for its ability to synthesize diverse and semantically meaningful images based on complex prompts.

    \item \textbf{VQDM~\cite{gu2022vector}.} A variant of diffusion models that combines vector quantization with diffusion processes.  VQDM performs image generation by discretizing the data distribution, which allows it to efficiently generate high-resolution images with more diversity. 

    \item \textbf{ADM~\cite{dhariwal2021diffusion}.}  A class of diffusion models that systematically removes components (e.g., attention mechanisms) to study their impact on generation quality. ADM is widely used for benchmarking due to its modular design and strong performance. 

    \item \textbf{Wukong~\cite{wukong2022}.}  A state-of-the-art text-to-image generation model trained on a massive dataset of Chinese text-image pairs. Wukong is particularly challenging for detection tasks due to its high-quality outputs and strong generalization capabilities. 
\end{itemize}

\section{Implementation Details}
\label{appendix_implement}
The feature extractor is a ResNet-50~\cite{he2016deep} trained on real and fake images generated by SD. 
The nearest neighbor reference set is built using the feature representations of 3,000 SD-generated images from training set. 
In experiments, we set the number of nearest neighbors \(k = 20\) for the KNN-based detection step. 
For baseline methods, we use their open-source implementations with recommended pretrained models and default hyperparameters to ensure a fair comparison.

\section{More Visualization Results}
\label{appendix_visualization}
To further validate the generalizability of PDA, we provide additional t-SNE visualizations and KNN distance distributions. 
These results are consistent with our findings from the main experiments and offer additional support for PDA’s alignment-based detection strategy.

\begin{figure}[htbp]
    \centering
      \includegraphics[width=0.46\linewidth]{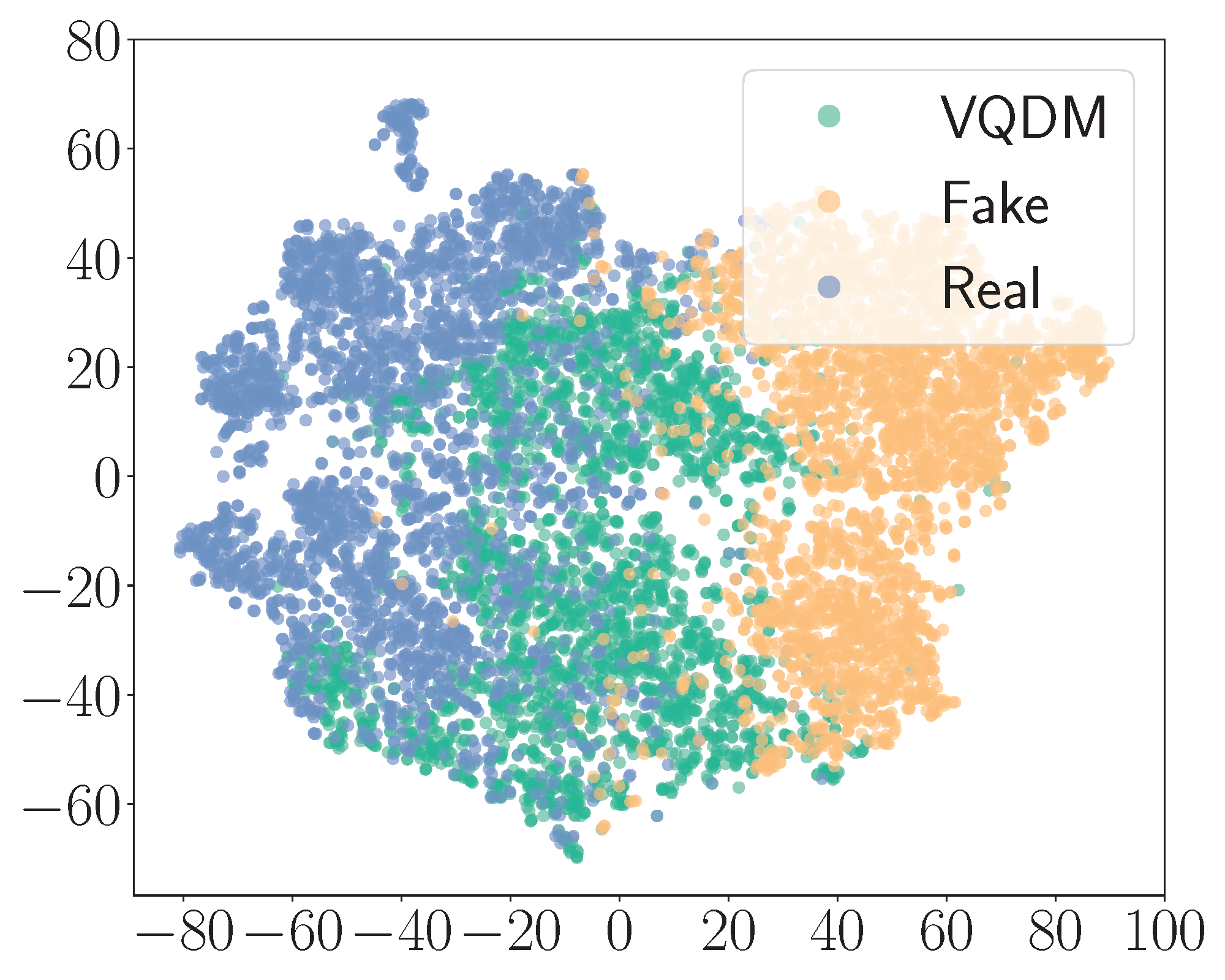}
    \includegraphics[width=0.46\linewidth]{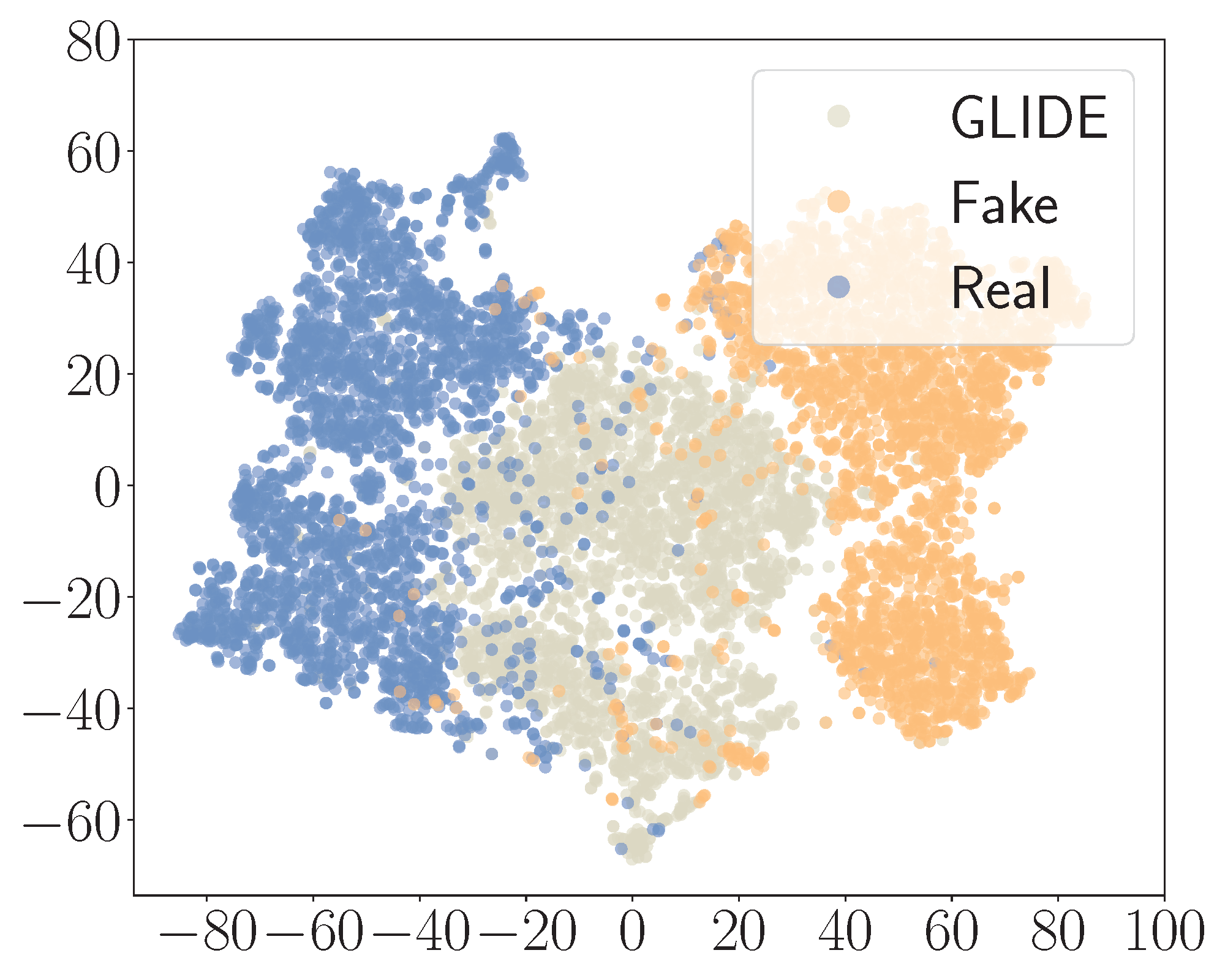}
  \\
    \includegraphics[width=0.46\linewidth]{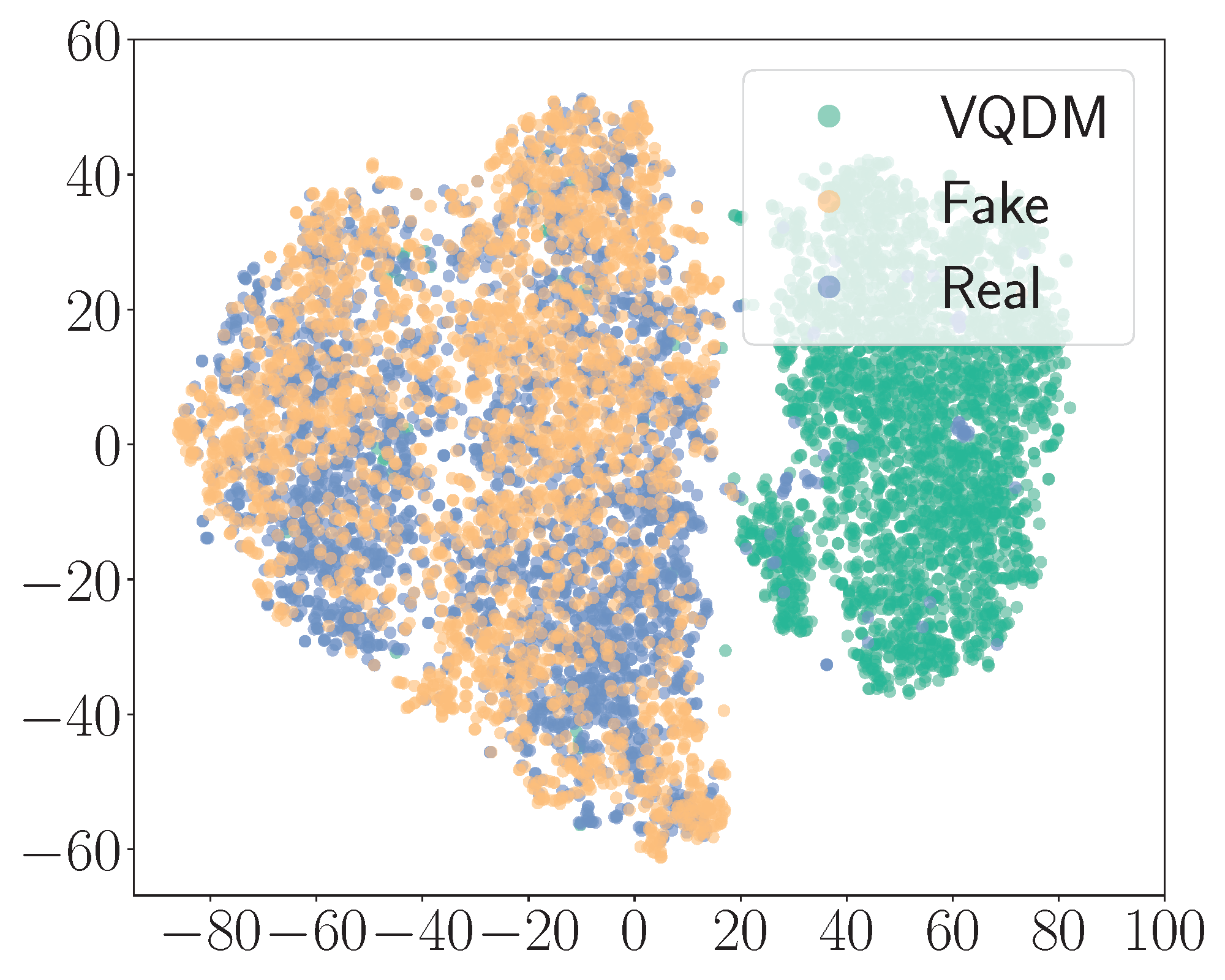}
    \includegraphics[width=0.46\linewidth]{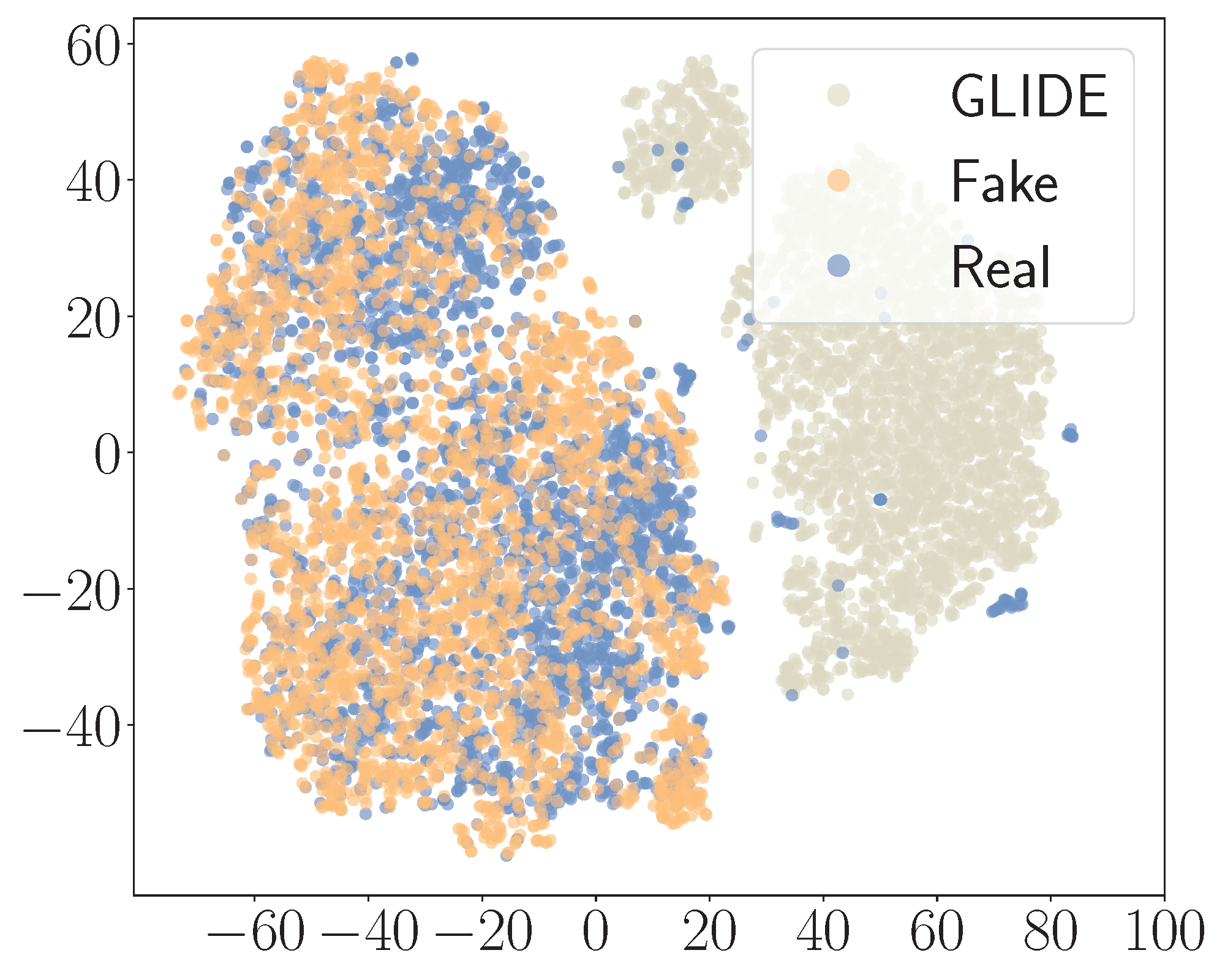}
\caption{T-SNE visualization. Rows correspond to feature spaces: original and regenerated images (“Fake” denotes the known fake distribution).}
\label{fig:T-SNE_more}
\end{figure}

As shown in \autoref{fig:T-SNE_more}, both GLIDE and VQDM-generated fake images exhibit substantial overlap with real images in the raw feature space, indicating that these fakes are hard to distinguish using standard classifiers. However, after regeneration through the known generator (SD), the pseudo-fake versions of real images become aligned with the known fake distribution, while regenerated unknown fakes from GLIDE and VQDM maintain distribution shifts.

\begin{figure}[htbp]
    \centering
    \includegraphics[width=0.46\linewidth]{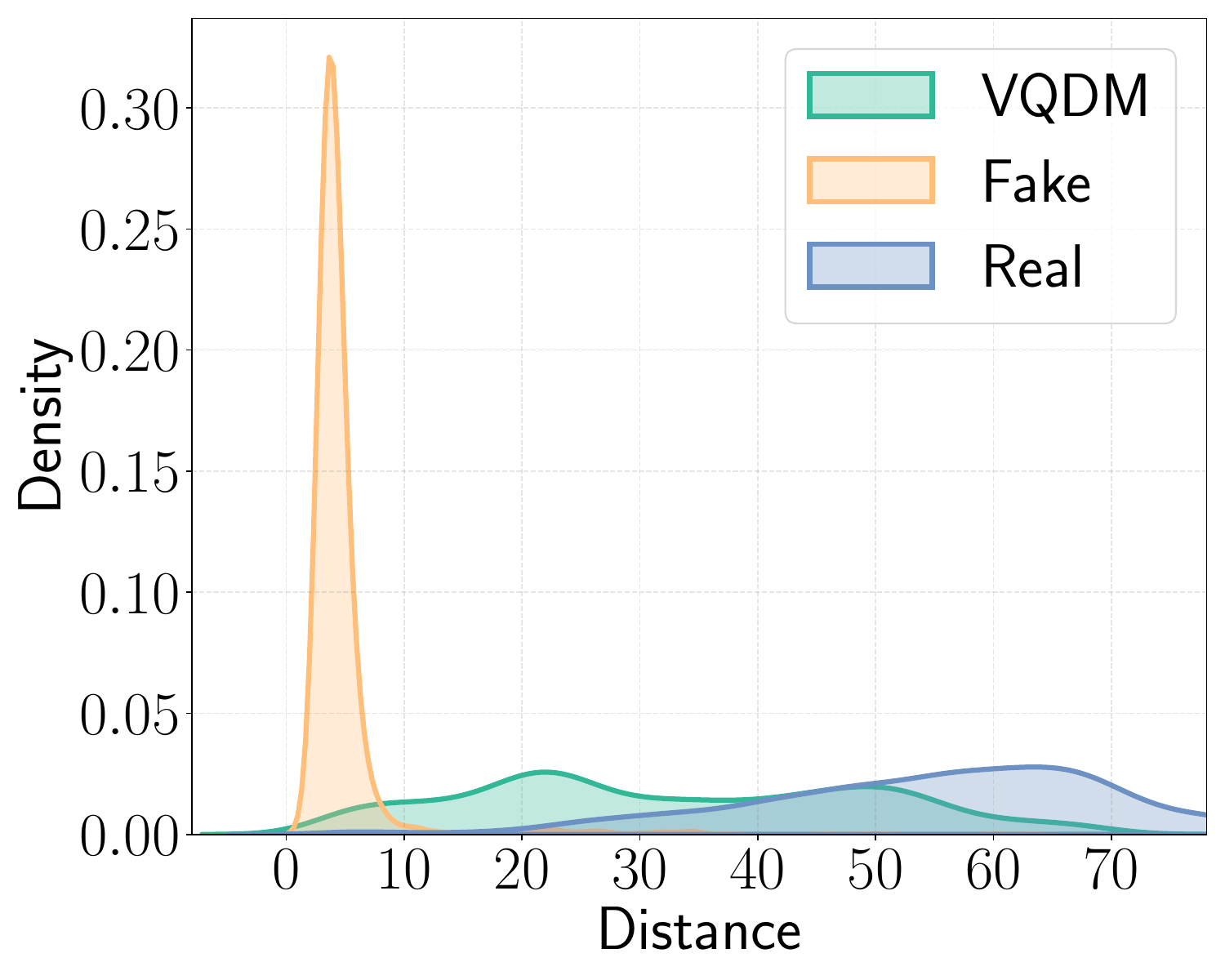} 
            \includegraphics[width=0.46\linewidth]{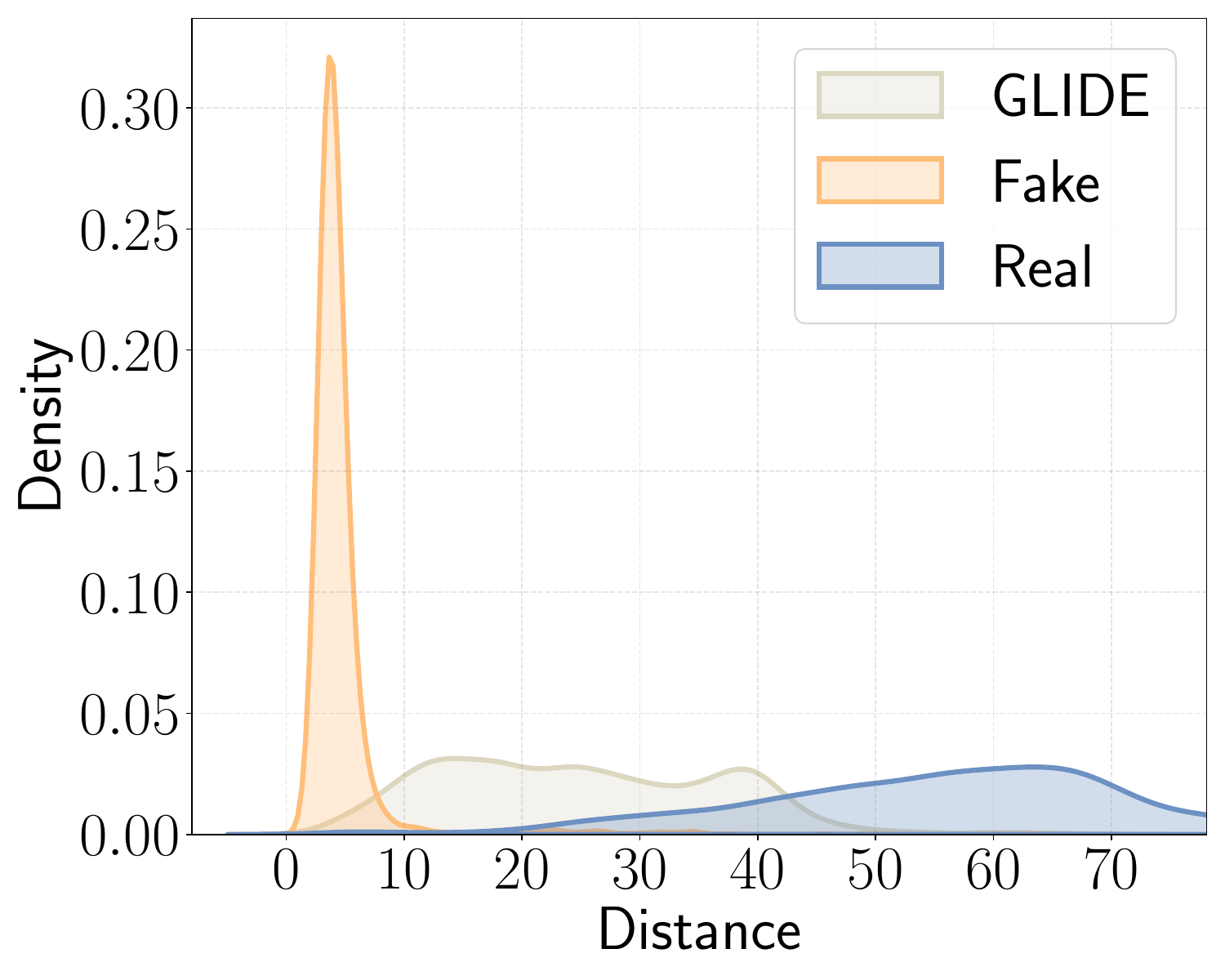} 
  \\
    \includegraphics[width=0.46\linewidth]{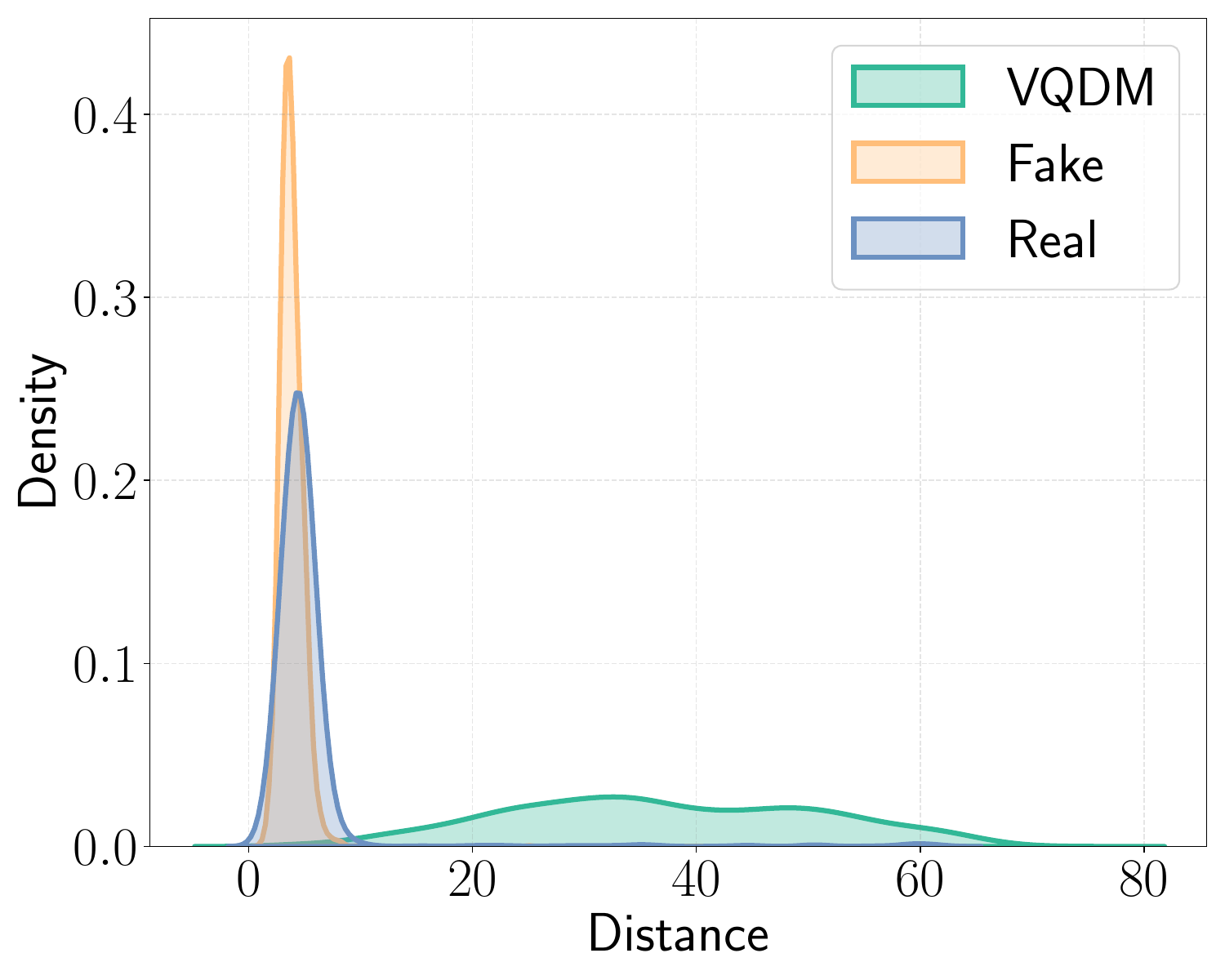} 
   \includegraphics[width=0.46\linewidth]{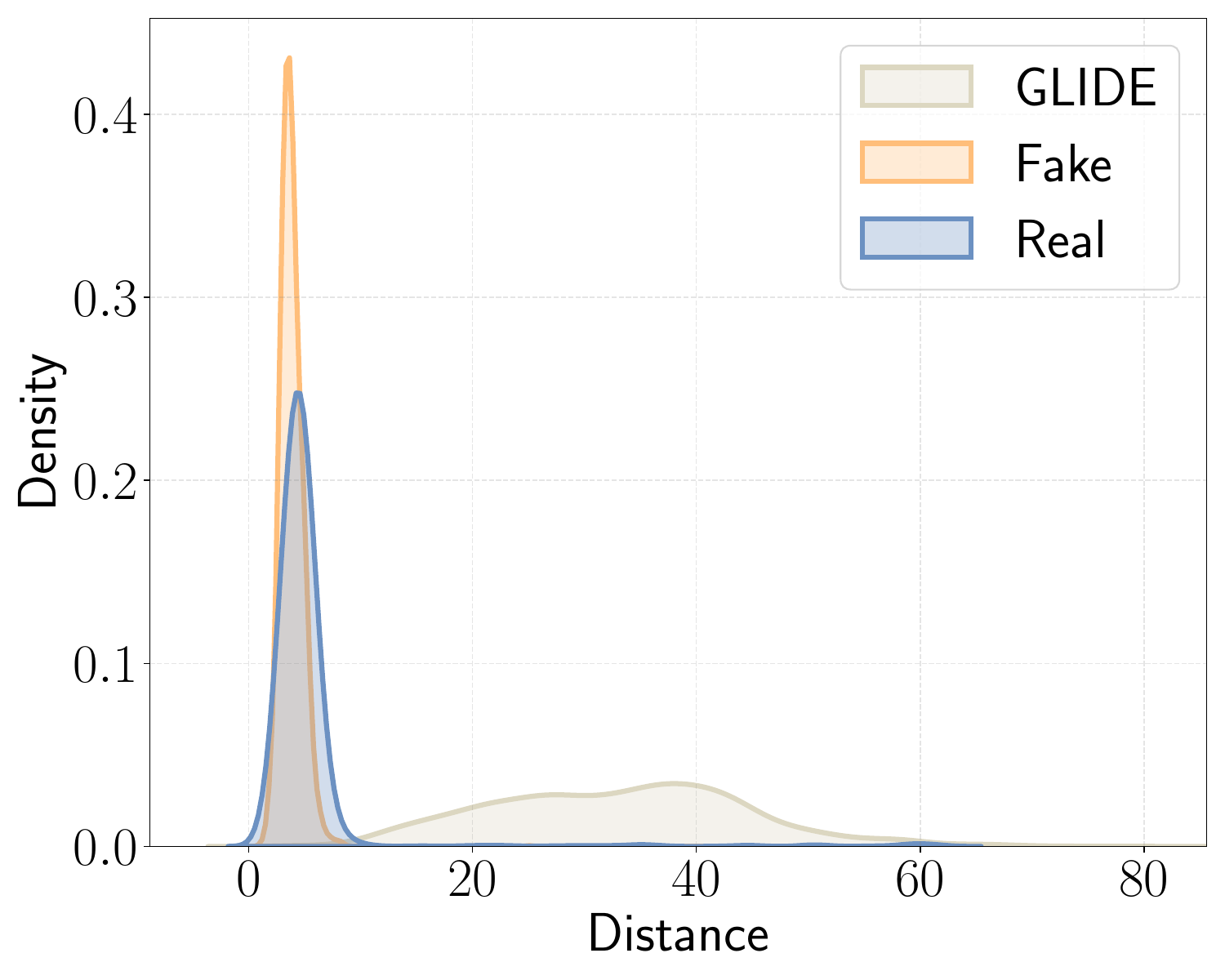} 
  \\
    \caption{KNN distance distributions. Rows correspond to feature spaces: original and regenerated images (“Fake” denotes the known fake distribution).}
\label{fig:KNN_more}
\end{figure}

As shown in \autoref{fig:KNN_more}, corresponding KNN distance plots further confirm this pattern. Pseudo-fakes derived from real inputs consistently show low KNN distances, while GLIDE and VQDM images exhibit persistently higher distances even after regeneration, due to their incompatible or mixed artifacts. These patterns reinforce the robustness of PDA in handling unknown generative models by leveraging post-hoc distribution alignment rather than relying on prior exposure to diverse fake distributions.

\end{document}